\documentclass[aps,prc,preprint,groupedaddress,nofootinbib,showpacs,preprintnumbers,amsmath,amssymb,superscriptaddress]{revtex4}

\usepackage{graphicx}
\usepackage{dcolumn}
\usepackage{bm}
\usepackage{amsmath}

\begin{document}

\def\R{\right} \def\L{\left} \def\Sp{\quad} \def\Sp2{\qquad}

\preprint{RCNP-Th02010}

\title{Path-integral hadronization for the nucleon and its interactions}

\author{L.J. Abu-Raddad} 
\email{laith@rcnp.osaka-u.ac.jp}
\affiliation{Research Center for Nuclear Physics, Osaka University, 10-1 Mihogaoka, Ibaraki,
Osaka 567-0047, Japan}  
\author{A. Hosaka} 
\email{hosaka@rcnp.osaka-u.ac.jp}
\affiliation{Research Center for Nuclear Physics, Osaka University, 10-1 Mihogaoka, Ibaraki,
Osaka 567-0047, Japan}  
\author{D. Ebert}
\email{debert@physik.hu-berlin.de}
\affiliation{Institut f\"ur Physik, Humboldt-Universit\"at zu Berlin, Invalidenstr.~110, 10115 Berlin,
Germany} 
\author{H. Toki}
\email{toki@rcnp.osaka-u.ac.jp}
\affiliation{Research Center for Nuclear Physics, Osaka University, 10-1 Mihogaoka, Ibaraki,
Osaka 567-0047, Japan}

\date{\today}

\begin{abstract}
Nucleon structure and the origin and nature of the nuclear force are
investigated in the context of a QCD-based effective field theory and
the path-integral method of hadronization. We start from a microscopic
model of quarks and diquarks where the gluons have been integrated
out.  In particular, we use the chiral Nambu-Jona-Lasinio model to
describe quark dynamics and assume that the nucleon can be conceived
as a quark-diquark relativistic bound state. The hadronization method
is then used to rewrite the problem in terms of the physical meson and
nucleon degrees of freedom.  Next, by employing a loop and derivative
expansion of the resulting quark/diquark determinants, we arrive at an
effective chiral meson-nucleon Lagrangian.  Nucleon properties such as
mass, coupling constants, electromagnetic radii, anomalous magnetic
moments, and form factors are derived using a theory of at most two
free parameters.
\end{abstract}


\maketitle
\section{Introduction \label{intro}}

The central problem in nuclear physics remains to understand the
origin and nature of the nuclear force. In spite of the belief that we
have attained the fundamental theory for the strong interactions--
Quantum chromodynamics (QCD), this theory still eludes a satisfactory
and complete description. The basic problem of QCD is that its natural
and fundamental degrees of freedom, quarks and gluons, are not the
observable baryon and meson states of the strong interaction. Thus
bridging the missing link between the fundamental and observable
degrees of freedom stands as one of the stark challenges of
nuclear/elementary particle physics today. Although we do have an ab
initio approach to solve this problem, that is lattice QCD, this
endeavor is still
miles away from achieving such a goal. This
naturally motivates us to resort to non-perturbative QCD-based approaches of
which this study is one.

In the present paper, we address this lingering missing link by
deriving a chiral meson-nucleon Lagrangian from a microscopic model of
quarks and diquarks using path-integral methods. Chiral symmetry and
its spontaneous breaking have consistently proven to be key concepts
in understanding meson and baryon structure and many features of the
nuclear force~\cite{Wein67,CWZ69,CCWZ69,EV81}.  The
gist
of this paper is as following: We start from a QCD-based effective field
theory to describe quark dynamics where the gluons have been
integrated out. This is the SU(2)$_L\; \times$ SU(2)$_R$ 
Nambu-Jona-Lasinio (NJL) model
that accommodates most of QCD
symmetries~\cite{NJL1,ER86}. Guided by general principles, we then
assume that the nucleon can be described as quark-diquark correlations
and introduce diquarks~\cite{IK66,Lich68} as elementary fields in
the problem. This assumption hinges upon the dynamical fact that two
quarks can combine to form a color anti-triplet leading with the third
quark to the formation of a color-singlet bound state, a baryon.
Moreover, this assertion is vindicated by a mounting experimental
evidence that diquarks play a dynamical role in hadrons~\cite{VW91,APEFL93,Ansel89,DonL80,Close81,FSJMLT82,KSS91,Stech87,Neu91}.

We verify that only two kinds of diquarks are relevant for nucleons:
the scalar isoscalar and the axial-vector isovector.  By introducing
composite meson and nucleon fields through the method of 
path-integral
hadronization and then using a loop and derivative expansion of the
resulting quark/diquark determinants, we arrive at an effective chiral
meson-nucleon Lagrangian. The path-integral hadronization
used here consists of two steps: bosonization to produce 
mesons as quark-antiquark
correlations and what we label as ``fermionization'' which generates
baryons as quark-diquark correlations. In our model, mass, coupling
constants, electromagnetic radii, anomalous magnetic moments, and form
factors of the composite nucleon are calculated in terms of at most two free
parameters.

In this fashion, our treatment parallels, in the sense of calculating
nucleon physical observables, the approach of using the Faddeev
equation~\cite{Fad61a,Fad61b} for three quark
states~\cite{ARWZ92,IBY93a,IBY93b,Meyer94}, or the approach of using
static quark exchange~\cite{HW95}, the Salpeter
equation~\cite{Salpeter52,Keiner96a,Keiner96b} or the fully
relativistic Bethe-Salpeter equation~\cite{SB51,HAOR97,OHAR98} for a
quark-diquark system. Nonetheless, our formalism yields, in addition
to nucleon observables, a Lagrangian of the quantum hadrodynamics (QHD)
type~\cite{Walecka74,SW86} that describes the rich meson-nucleon
interactions in a fully covariant and chirally symmetric formalism.

While this program is applied to the case of deriving an effective
Lagrangian for nucleons and mesons, it is certainly of general nature
and can possibly be applied alternatively to yield prolifically other
baryons and their interactions such
as the $\Delta$ particle. Moreover, the idea of using path-integral
techniques to transform a Lagrangian from its fundamental to its
composite degrees of freedom is a powerful concept in physics of
immense impact and utility. As a matter of fact, the authors of
Ref.~\cite{IMO01} have recently invoked such path-integral techniques
in their study of high-temperature superconductivity.  They succeeded
in doing so by converting a model of strongly-correlated electrons
into an effective U(1) gauge field theory in terms of composite
fields.

The use of path-integral hadronization to derive a meson-baryon
Lagrangian has been introduced in Ref.~\cite{Cahill89, Rein90, Ebka92}
and applied to baryons with heavy quarks~\cite{EFKR96,EFKR98}.  Based
on these ideas, the authors of Ref.~\cite{EJ98} attempted to construct
such an effective Lagrangian for the nucleon using only scalar
diquarks. They derived correctly the structure of the meson-nucleon
Lagrangian, proved the Goldberger-Treiman relation and attempted to
evaluate the axial-vector coupling constant $g_A$ as an application of
their formalism.  Their analysis and numerics for $g_A$ contain,
however, few problems as well as an uncertainty due to the lack of a
proper gauge-invariant regularization scheme. In the present paper, we
extend their work by deriving the structure of the corresponding
Lagrangian using both axial-vector and scalar diquarks, employ a
gauge-invariant regularization scheme throughout our analysis, and
verify the Ward-Takahashi identity and the Goldberger-Treiman relation.
Furthermore, we present a full numerical study of various nucleon
observables for the case of scalar diquarks drawing special attention
to the role of an intrinsic diquark form factor. We concentrate our
analysis first on the scalar-diquark case for simplicity and due to
the predicted dominance of this type of diquark in the
nucleon~\cite{CRCP87,Rein90,Vogl90,MBIY02}.  Thus after
more than ten years since the introduction of the idea of path-integral
hadronization, this formalism is finally used to speak itself in
calculating nucleon structure and its observables.

The paper has been organized as follows. In Sec.~\ref{micromodel} a
microscopic model for quarks, diquarks, and their interactions is
developed and meson and nucleon fields are introduced as auxiliary
fields in the problem.  The hadronization method is then invoked in
Sec.~\ref{derivation} to rewrite the microscopic Lagrangian in terms
of composite meson and nucleon fields. Next, a loop and derivative
expansion is employed to calculate several terms in the Lagrangian
including the nucleon self-energy and electromagnetic vertex. The issue of
regularization is also examined and the
Ward-Takahashi identity and the Goldberger-Treiman relation are
verified. In Sec.~\ref{num} a full numerical study for the nucleon is
presented. Finally a summary and conclusions are provided in
Sec.~\ref{concl} as well as a discussion of some of the challenges and
opportunities that remain.

\section{A microscopic model of quarks, diquarks and their interactions}
\label{micromodel}

\subsection{
Nambu-Jona-Lasinio Model}
 
In our model we treat the quarks using the NJL model which is a
successful effective field theory where quarks interact through a
four-point local fermion-fermion coupling. The highlights of the model
are its incorporation of all global symmetries of QCD as well as its
prediction of many features of QCD such as dynamical chiral
symmetry breaking and its restoration
~\cite{ER86,VW91,HK94,ERV94}. Moreover, this
model has been motivated, if also not derived, using lattice
QCD~\cite{KS81}, continuum QCD
~\cite{HK94,ERV94,And87,Ball90,Rein91,BBR93},
and Yang-Mills theories~\cite{SRAL}. The locality assumption of the
model has been justified for low energy QCD ~\cite{SRAL}, and inspired
by strong-coupling lattice quantum electrodynamics
(QED)~\cite{Gock90,VW91}. The main problem of the NJL model continues
to be the absence of confinement. Therefore, the success of the model
rests on observables that are insensitive to the details of
confinement. It is noteworthy here that there exists various attempts to include the effects of
confinement within the NJL model~\cite{BEE93,HK94}.

We start from an NJL Lagrangian satisfying SU(2)$_L\; \times$
SU(2)$_R$ chiral symmetry:
\begin{equation}
{\cal L_{\text{NJL}}} = {\bar q} (i\rlap/\partial -m_0) q + \frac{G}{2} \left[
({\bar q} q)^2 + ({\bar q} i \gamma_5 \vec{\tau} q)^2 \right]\;.
\end{equation}
Here $q$ is the current quark field, $\vec{\tau}$ are the isospin
(flavor) Pauli matrices, $G$
is the NJL coupling constant, and $m_0$ is the current quark mass
which explicitly breaks chiral symmetry. The color and flavor indices
are suppressed in this expression and assumed to be so for the rest of
the paper unless explicitly shown. Starting from this Lagrangian, we construct the corresponding vacuum partition function as
\begin{eqnarray}
{\cal Z} = {\mathcal N_1} \int {\mathcal D} q {\mathcal D} \bar{q}\; {\exp}\;
i\int {\rm d}^4x \; \L[{\cal L_{\text{NJL}}}\R]\;,
\label{znjl}
\end{eqnarray}
where ${\mathcal N_1}$ is a normalization constant.

\subsection{Introduction of meson fields}

Composite scalar ($\sigma \sim \bar q q$) and
pseudoscalar ($\vec{\pi} \sim \bar q i \gamma_5 \vec \tau q$) meson fields are introduced as auxiliary fields in the
problem. This is done by multiplying the NJL partition function of
Eq.~(\ref{znjl}) by the term (with ${\mathcal N_2}$ being another normalization constant)
\begin{eqnarray}
{\mathcal N_2} \int {\mathcal D} \sigma {\mathcal D} \vec\pi\; {\exp}\;
i\int {\rm d}^4x \; \L[-\frac{1}{2G}(\sigma^2+\vec\pi^2)\R]\;.
\label{mesaux}
\end{eqnarray}
At this stage, no modifications have been made to the 
underlying dynamics of the
Lagrangian as this multiplicative factor is merely an overall constant in
the partition function. We impose the following transformation:
\begin{eqnarray}
\sigma &\longrightarrow & \sigma + G \bar{q} q\;,\nonumber\\
\pi^i &\longrightarrow & \pi^i + G \bar{q} i \gamma^5 \tau^i q\;,
\label{mestrans}
\end{eqnarray}
in order to eliminate the 
quadratic terms ($\sim
(\bar{q} q)^2$) of the NJL Lagrangian. 
Using translational invariance of the integration measure 
${\mathcal D} \sigma {\mathcal D} \vec\pi $,
this results in the expression:
\begin{eqnarray}
{\mathcal N_3} \int {\mathcal D} \sigma {\mathcal D} \vec\pi
{\mathcal D} q {\mathcal D} \bar{q} \; {\exp}\;
{i\int {\rm d}^4x \; \L[\bar{q}\L(i\rlap/\partial - m_0 
- \sigma - i\gamma_5 \vec\tau\cdot\vec\pi\R) q
-\frac{1}{2G}(\sigma^2+\vec\pi^2)\R]}\;.
\label{sembos}
\end{eqnarray}
The prescribed change in field 
variables
is nothing but the
Hubbard-Stratonovich transformation~\cite{Strat58,Hubbard59}. We label the
resulting Lagrangian as the ``semi-bosonized'' one since we
have already introduced the boson (meson) fields but have not yet
integrated over the quark ones.
The current quark mass $m_0$ is then absorbed into the definition of the
field $\sigma$ and the meson fields are further transformed according to the non-linear parameterization $\L[\sigma,\pi\R]
\rightarrow \L[\sigma^\prime,\Phi\R]$ 
\begin{eqnarray}
\sigma+i\gamma_5\vec\tau\cdot\vec\pi&= &\left( m_q +
\sigma^\prime\right){\exp}{\L(-\frac{i}{F_\pi}\gamma_5\vec\tau\cdot\vec\Phi\R)}\;,
\label{sipitrans}
\end{eqnarray}
where $F_\pi = 93$~MeV is the pion decay constant and $m_q\equiv
\langle\sigma\rangle_0$ is the constituent quark mass which is fixed
through a gap equation in the meson sector
~\cite{ER86,HK94,ERV94}. 
Accordingly, the NJL
Lagrangian is converted to 
\begin{equation}
{\cal L_{\text{NJL}}} =  \delta{\cal L}_{\rm sb} \;-\; \frac{1}{2G}\L(\sigma^{\prime}+m_q\R)^2 \;+\;
{\bar q} \L[ i\rlap/\partial - \left( m_q +
\sigma^\prime\right) {\exp}\L({-\frac{\rm
i}{F_\pi}\gamma_5\vec\tau\cdot\vec\Phi}\R)\R] q\;. 
\label{sembos2}
\end{equation}
Here, the $\delta{\cal L}_{\rm sb} = \mathcal O(m_0)$ is the
symmetry-breaking mass term given by
\begin{eqnarray}
\delta{\cal L}_{\rm sb} =
\frac{m_q+\sigma^\prime}{16\kern.5mmG}m_0\kern1mm {\rm
tr}\left[{\exp}\L({-\frac{i}{F_\pi}
\gamma_5{\vec\tau\cdot\vec\Phi}}\R)+ {\rm h.c.}  \right]\;,
\end{eqnarray}
where the trace is taken over flavor and Dirac 
indices.

\subsection{Diquarks}

In studying the Lorentz and flavor structure of the $qq$ correlations,
we find five possible types: the scalar $\tilde{q} q$, pseudo-scalar
$\tilde{q} i \gamma_5 q$, vector $\tilde{q} \gamma_\mu q$,
axial-vector $\tilde{q} \gamma_\mu \gamma_5 q$, and tensor $\tilde{q}
\sigma_{\mu \nu} q$ diquarks. Here $\tilde{q} \equiv q^{\rm T}
C^{-1}\gamma_5 i \tau_2$ where $\text{T}$ stands for transpose and
$C^{-1} = i \gamma_2 \gamma_0$ is the inverse charge-conjugation
matrix. Moreover, we identify two isospin structures for each of these
five Lorentz $qq$ formations. Explicitly, we have an isoscalar and
isovector diquarks by inserting $\openone$ and $\vec{\tau}$ between
$\tilde{q}$ and ${q}$. 
It has to be noted here that the $\tilde{q}$ spinor
has the same transformation properties as $\bar{q}$ in the Lorentz and
isospin groups.

A question arises concerning how many of these ten diquarks are needed
to form the nucleon. Using permutation symmetry and Fierz
transformation, we have verified an earlier assertion~\cite{EPT83}
that only two diquark formations are independent for the nucleon if
the nucleon field is to be written as a local operator of three quarks.
This result is consistent with the fact that in constituent quark
models~\cite{Greiner94}, the nucleon wave function is constructed
using only scalar and axial-vector diquarks. Hence, we introduce two
diquarks as elementary complex fields: ${\vec D}^{\mu}$ as
an axial-vector isovector field with electric charge = \{4/3, 1/3,
-2/3\} and $D$ as a scalar isoscalar field with a charge = \{1/3\}.

\subsection{Quark-diquark interaction terms}

Quark-diquark interaction terms are introduced to form the nucleon
as a relativistic bound state of quarks and diquarks. We consider such an interaction in a local form. This
is essentially the static approximation of solving the three-body
equations for baryons within the NJL model~\cite{IBY93b,MBIY02}.
It is more convenient
here, in terms of forming a chirally invariant quark-diquark couplings,
to work with the chirally rotated 
``constituent'' quark field $\chi$
defined by~\footnote{The chiral rotation of Eq.~(\ref{consq}) induces anomalous terms as
the phase of the integral measure. In this paper, we are
concerned only with the non-anomalous processes.} 
\begin{equation}
\chi \equiv \exp\L({-\frac{i}{F_\pi} \gamma^5
\frac{\vec{\tau}}{2} \cdot \vec{\Phi}}\R) q\;.
\label{consq}
\end{equation}

The range of possible
symmetry preserving interaction terms is limited~\footnote{If we work with the current quark field $q$,
that is in the linear representation of chiral symmetry, a
vector-isoscalar diquark is necessary as a chiral partner of the
axial-vector-isovector one.}. This provides a
highly welcomed dynamical constraint in our treatment.
Discarding for a moment interaction terms describing a
possible scalar-axial-vector mixing 
(see Eq.~(\ref{lsemibos}) below),
we may choose the following term 
for the quark-scalar-diquark interaction:
\begin{equation}
{\cal{L}}_{q D} \sim \tilde{G} \; \L(\bar{\chi} D^\dag\R)
\L(D \chi\R)\;,
\end{equation}
while we may select
\begin{equation}
{\cal{L}}_{q {\vec D}^{\mu}} \sim \tilde{G} \;
 (\bar{\chi}\gamma^\mu \gamma^5\:\vec{\tau}\cdot{\vec
 D}^\dag_{\mu})\; ({\vec D}_{\nu}\cdot\vec{\tau}\:\gamma^\nu \gamma^5
 \chi)\;,
\end{equation}
for the quark-axial-vector-diquark coupling. Our choice for the full interaction term is dictated by the need to
produce the nucleon as a linear combination of axial-vector and scalar
diquarks according to
\begin{equation}
B\;\; \sim \tilde {G} \; \L(\sin{\theta} \;{\vec
D}_{\mu}\cdot\vec{\tau}\;\gamma^\mu \gamma^5\; \chi \;+\;
\cos{\theta}\; D \;\chi\R)\;.
\end{equation}
In the above expressions $\tilde{G}$ is the quark-diquark coupling
constant 
with mass dimension $[\tilde {G}]=m^{-1}$,
while $\theta$ is a mixing angle for the two
diquark contributions. These are, as we shall discuss below (see
Sub-Sec.~\ref{param}), the only free parameters in our model.

\subsection{Microscopic Lagrangian}

Electromagnetic interactions are introduced in our model through the
canonical method of covariant derivatives in the quark and diquark
Lagrangians. We proceed to form the microscopic Lagrangian by 
batching the diquark contributions, the
quark-diquark interaction terms including mixing, and the
semi-bosonized NJL Lagrangian of Eq.~(\ref{sembos2}) after the field
transformations of Eq.~(\ref{consq}), to obtain the following
Lagrangian as our input model:
\begin{eqnarray}
{\cal L} &=& \bar{\chi}\; S^{-1} \; \chi \;-\;
\frac{1}{2G}(\sigma^{\prime}+m_q)^2\;+
\; \delta{\cal L}_{\rm sb}
\;+\; D^\dag \;\Delta^{-1} \;D \;+\; {\vec{D}^{\dag\;\mu}} \;
{\tilde{\Delta}}^{-1}_{\mu \nu} \; \vec{D}^{\nu} \;+\; \nonumber\\ &&
\tilde{G} \L( \sin{\theta} \; \bar{\chi}\gamma^\mu \gamma^5
\:\vec{\tau}\cdot {\vec D}^\dag_{\mu} \;+\; \cos{\theta}
\;\bar{\chi} D^\dag\R)\; \L(\sin{\theta}\; {\vec D}_{\nu}\cdot
\vec{\tau} \: \gamma^\nu \gamma^5 \chi \;+\; \cos{\theta} \;D \chi \R)\;,
\label{lsemibos}
\end{eqnarray}
where
\begin{subequations}
\begin{eqnarray}
S^{-1} &=& S_0^{-1} \;+\; {\cal M}\;,\\
{\cal M} &=& -\left[ \gamma^\mu \frac{\vec{\tau}}{2}\cdot \vec{\mathcal
V}^{\pi}_\mu \;+\; \gamma^\mu \gamma^5 \frac{\vec{\tau}}{2}\cdot
\vec{\mathcal A}^\pi_\mu \;+\; \sigma^\prime \;+\; \gamma^\mu Q_q
A^{\rm EM}_{\mu} \right]\;.
\end{eqnarray}
\end{subequations}
Here $S^{-1}$ is the modified inverse propagator that includes the free inverse
propagator $S_0^{-1} = (i\rlap/\partial - m_q)$ for the quark field
and the interaction matrix ${\cal M}$. Note that 
$Q_q = \text{diag}\L( 2/3, -1/3\R)$ is the quark charge.
The ${\cal M}$ matrix contains all interaction vertices of the quark with
meson and electromagnetic fields. 
The quark interacts with the pion
through vector $\vec{\mathcal V}^{\pi}_\mu$ and axial-vector
$\vec{\mathcal A}^\pi_\mu $ functions of the pion field. These
arise from the derivative term $i\rlap/\partial$ following the
transformation of Eq.~(\ref{consq}). Precisely,
these functions are
defined through the Cartan
decomposition ($\vec{\xi} \equiv
\frac{\vec\Phi}{F_\pi}$):
\begin{eqnarray}
{\exp}\L({-\frac{i}{2}\gamma^5{\vec\tau\cdot\vec\xi}}\R)
\;\partial_\mu\ {\exp}\L({\frac{i}{2}
\gamma^5{\vec\tau\cdot\vec\xi}}\R) \;=\; {\textstyle\frac{i}{2}\;
\gamma^5\;{\vec\tau} }\cdot\vec {\mathcal A}^\pi_\mu(\xi)\;+\;
{\textstyle\frac{i}{2} \; {\vec\tau}}\cdot\vec{\mathcal
V}^\pi_\mu(\xi)\;.
\label{neun}
\end{eqnarray}
The ${\cal M}$ matrix also encompass pion-photon and weak gauge boson
vertices which are not shown for brevity.

In the above Lagrangian, we use the modified scalar diquark inverse propagator
\begin{eqnarray}
\Delta^{-1} = \Delta_{0}^{-1} + i Q_S A^{\rm EM}_{\mu}
(\overrightarrow{\partial^\mu} - \overleftarrow{\partial^\mu})\;,
\label{scalarD}
\end{eqnarray}
and the modified axial-vector diquark inverse propagator 
\begin{eqnarray}
{\tilde{\Delta}}^{-1}_{\mu \nu} = {{\tilde{\Delta}}_0\;}^{-1}_{\mu
\nu} \;+\; i Q_A \left[ (A^{\rm EM}_{\mu} \overleftarrow{\partial_\nu}
- A^{\rm EM}_{\nu} \overrightarrow{\partial_\mu} ) \;-\; g_{\mu \nu}
A^{\rm EM}_{\alpha}(\overleftarrow{\partial^\alpha} -
\overrightarrow{\partial^\alpha}) \right]\;,
\label{axvecD}
\end{eqnarray}
where we have omitted ${\cal O}(Q_S^2)$ and ${\cal O}(Q_A^2)$ terms
for brevity.
Each of these expressions includes the free inverse propagator as the kinetic
and mass term, and the electromagnetic interaction vertex. Explicitly,
the free inverse propagator for the scalar diquark is
 \begin{eqnarray}
\Delta_{0}^{-1} = -(\partial^2  + M_S^2)\;,
\end{eqnarray}
while the one for the axial-vector diquark is
 \begin{eqnarray}
{\tilde{\Delta}}^{-1}_{\mu \nu} =  \openone\otimes
[g_{\mu \nu} (\partial^2  + M_A^2) - \partial_\mu
\partial_\nu]\;,
\end{eqnarray}
where $\openone$ is the unit matrix in the
three-dimensional isospace.  
Here $Q_S = 1/3$ and $Q_A = \text{diag}\L(4/3,1/3,-2/3\R)$ are
the charges of the scalar and axial-vector diquarks respectively. The
modified axial-vector diquark propagator  
encompasses (not shown) weak interaction terms.

\subsection{Introduction of nucleon fields}

There is still one missing component in our microscopic model:
collective nucleon fields. Therefore, a nucleon field $B$ is introduced as an
auxiliary one by multiplying the partition function of the Lagrangian
of Eq.~(\ref{lsemibos}) by the term
\begin{equation}
{\mathcal N_4} \int {\mathcal D} B {\mathcal D} \bar{B}\; {\exp}\;
i\int {\rm d}^4x \; \L[-\frac{1}{\tilde{G}}\bar{B} B\R]\;,
\end{equation}
where ${\mathcal N_4}$ is a normalization constant. In a similar fashion to Eq.~(\ref{mesaux}), (\ref{mestrans}) and (\ref{sembos}), we transform the
field configuration according to:
\begin{eqnarray}
B &\longrightarrow & B + \tilde{G}\L(\sin{\theta}\; {\vec
D}_{\nu}\cdot \vec{\tau} \: \gamma^\nu \gamma^5 \chi \;+\;
\cos{\theta} \;D \chi \R)\;, \nonumber\\
\bar{B} &\longrightarrow & \bar{B} + \tilde{G}\L( \sin{\theta} \;
\bar{\chi}\gamma^\mu \gamma^5 \:\vec{\tau}\cdot {\vec D}^\dag_{\mu}
\;+\; \cos{\theta} \;\bar{\chi} D^\dag\R)\;.
\end{eqnarray}
As a result, the quark-diquark interaction term in
Eq.~(\ref{lsemibos}) is rewritten as:  
\begin{eqnarray}
\exp\; i\int {\rm d}^4x\; \tilde{G} \L( \sin{\theta} \;
\bar{\chi}\gamma^\mu \gamma^5 \:\vec{\tau}\cdot {\vec D}^\dag_{\mu}
\;+\; \cos{\theta} \;\bar{\chi} D^\dag\R)\; \L(\sin{\theta}\; {\vec
D}_{\nu}\cdot \vec{\tau} \: \gamma^\nu \gamma^5 \chi \;+\;
\cos{\theta} \;D \chi \R) \nonumber\\ = \;\mathcal N_4\int
{\mathcal D} B{\mathcal D} \bar B\; \exp \;\bigg\{\; i \int {\rm
d}^4x\;\bigg[\;\frac{-1}{\tilde G}\;\bar BB \;-\;
\bar{B}\;\L(\sin{\theta}\; {\vec D}_{\nu}\cdot \vec{\tau} \:
\gamma^\nu \gamma^5 \chi \;+\; \cos{\theta} \;D \chi \R) \nonumber\\
-\; \L( \sin{\theta} \; \bar{\chi}\gamma^\mu \gamma^5
\:\vec{\tau}\cdot {\vec D}^\dag_{\mu} \;+\; \cos{\theta}
\;\bar{\chi} D^\dag\R)\; B\;\bigg]\;\bigg\}\;.
\label{hadronization}
\end{eqnarray}
This procedure completes the introduction of composite meson and
nucleon fields into the problem and wrap up 
the construction of the microscopic model.

\section{Derivation of a meson-nucleon Lagrangian}
\label{derivation}

\subsection{Hadronization of the microscopic model}

At this point we have a Lagrangian that involves only quarks and
diquarks as dynamical fields with kinetic and mass terms while the meson and
nucleon fields are merely auxiliary ones. Additionally, the quark and
diquark fields appear in bilinear forms appropriate for integration as
a consequence 
of eliminating the interaction terms.  Thus, we
rearrange the expressions involving the quark fields into the form
$\bar{\chi}\; S^{-1} \chi - \bar{\eta} \chi - \bar{\chi}\eta$, and use
the fermion path-integral identity (Det stands for determinant):
\begin{equation}
\int {\mathcal D} \chi {\mathcal D} \bar{\chi}\; {\exp}\; i \;\int\; \L(
\bar{\chi} S^{-1} \chi - \bar{\eta} \chi
 - \bar{\chi}\eta \R) = \text{Det}\L(S^{-1}\R)
\exp\L( -i \;\int  \; \bar{\eta}\: S \:\eta\R)\;,
\end{equation}
to integrate over the quark fields. In doing so, we would have
accomplished the path-integral
bosonization that delivers to mesons their full dynamical
character~\cite{ER86,ERV94}. We still need to integrate over the
axial-vector and scalar diquark fields in order to
achieve a
meson-nucleon Lagrangian.  Thus, we cast the terms that involve the
diquark fields into the form $ \varphi^\dag {\cal K} \varphi$ where
$\varphi = \L( \vec{D}^\mu, D\R)^{\rm T}$, and use the boson path-integral
identity:
\begin{equation}
\int {\mathcal D} \varphi^\dag {\mathcal D} \varphi \; {\exp}\; i \int \; \L( \varphi {\cal K} \varphi^\dag\R) =
\L[\text{Det}({\cal K})\R]^{-1}\;.
\end{equation}
This final
integration procedure is what we label as fermionization as 
it produces fermions from boson-fermion correlations. The
quark-diquark dynamics has
been absorbed by the composite
meson and nucleon fields. We have at last fully
``hadronized'' the quark and diquark Lagrangian.  The microscopic model of
quarks and diquarks has been converted into a ``macroscopic'' model of
mesons and nucleons possessing the same (approximate) chiral
symmetry as the original microscopic fields.  Notice that the quarks and
diquarks do now appear 
only as virtual particles in loops and are
described by corresponding propagators and interaction vertices.

Next, we use the relation $\text{Det}(M) = \exp \text{tr} \ln
\L(M\R)$, to rewrite the determinants as 
Lagrangian terms of meson and nucleon fields. Thereupon, we arrive
at a compact chiral meson-nucleon Lagrangian given by
\begin{eqnarray}
{\cal L}_{\rm eff} &=& \delta{\cal L}_{\rm sb} \;-\;
\frac{1}{2G}(\sigma^{\prime}+m_q)^2 \;-\; i \;{\rm tr\; ln} S^{-1}
\;-\; \frac{1}{\tilde{G}}\; \bar{B} B \;+\; i\;{\rm tr \; ln} ( 1
\;-\; \Box ) \;+\; \nonumber \\ &&i\; {\rm \;tr\; ln} ( 1 \;-\;
\Delta_0\; {\rm EM}\;{\rm Int}) \;+\; i\;{\rm tr\; ln} ( 1 \;-\;
\tilde{\Delta}_0 \;{\rm EM}\;{\rm Int})\;.
\label{effL}
\end{eqnarray}
Here the trace is over color, flavor, and Lorentz indices while the
``EM Int'' label stands for the electromagnetic interaction terms of
each of the diquarks as given in Eq.~(\ref{scalarD}) and
Eq.~(\ref{axvecD}). Furthermore,
\begin{subequations}
\begin{eqnarray}
\Box & = & \begin{pmatrix} {\mathcal{A}}& {{\cal{F}}_2}\\
 {{\cal{F}}_1} &{\mathcal S}
\end{pmatrix},
\end{eqnarray}
\text{where}
\begin{eqnarray}
{\mathcal A}^{\mu i,\,\nu j} & = & {\sin}^2\theta \
\bar{B} \; 
\gamma_\rho \gamma^5\;
  {\tau}_{k} \;\tilde{\Delta}^{\rho k,\,\mu i} \;
 S\; {\tau}^{j}\;
  \gamma^\nu \gamma^5\; B\;,\\ 
{\mathcal S}& = & {\cos}^2\theta\; \bar{B}\;\Delta
  \;S\;B\;,\;\; \;\;\;\;\\ 
({{\mathcal F}}_1)^{\nu j}& = & \sin{\theta} \cos{\theta}\;\bar{B}\; 
\Delta \;S\; 
 {\tau^j}\;\gamma^\nu \gamma^5\;  B\;,\\
({{\mathcal F}}_2)^{\mu i}& = & \sin{\theta} \cos{\theta} \;
\;\bar{B} \;\tilde\Delta^{\rho k,\, \mu i}\;\gamma_\rho \gamma^5\; {\tau_k} \; S\; B\;.
\end{eqnarray}
\end{subequations}
The configuration and color 
indices have been omitted for simplicity.
The effective hadron Lagrangian of Eq.~(\ref{effL}) contains plenty of
rich physics. It encompasses, through the loop and derivative
expansion in the ${\rm tr\; ln}$
terms, kinetic and mass terms for nucleons
and mesons together with a multitude of possible interaction terms of mesons,
nucleons, and electroweak gauge bosons. It comprises terms describing
the various electroweak interactions of mesons and nucleons such as
meson photoproduction (the Kroll-Ruderman terms~\cite{Hosaka01}), and
in addition, it includes terms for meson-meson and nucleon-nucleon
scattering. Nonetheless, the most desired part of the Lagrangian is
the prized meson-nucleon interaction and nucleon-nucleon vertices
which delineate the nuclear force.

\subsection{Self-energy diagram and kinetic terms}

The physics in the Lagrangian becomes manifest in terms of loop and derivative expansions of the
resulting quark-diquark determinants. We concentrate here on the
nucleon sector which is contained in the terms $i\;{\rm tr\; \rm ln} (
1 - \Box ) - \frac{1}{\tilde{G}}\; \bar{B} B$. Nicely, the expansion
\begin{eqnarray}
{\rm tr\; \rm ln} ( 1 - \Box ) = -\rm tr \L( \Box + \frac{\Box^2}{2} + \frac{\Box^3}{3} +
\cdots\;\R),
\label{exp1}
\end{eqnarray}
is an expansion in the number of nucleon fields.  Moreover, since
\begin{eqnarray}
S = (1 \;+\; S_0 {\mathcal M})^{-1} S_0 = S_0 - S_0 \:{\mathcal M}\:
S_0 \;+\; S_0 \: {\mathcal M} \: S_0\: {\mathcal M}\: S_0 \;+\;
\cdots\;,
\label{exp2}
\end{eqnarray}
each term in the logarithmic series leads to an expansion in terms of
  the number of interaction vertices : $0, 1, 2, 3, \cdots$. 

\begin{figure}
\includegraphics{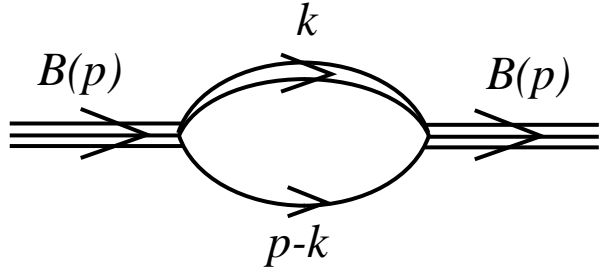}
\caption{\label{fig1} The Feynman diagram for the nucleon
self-energy which generates the nucleon kinetic and mass terms and produces the mass equation that determines the nucleon
mass.}
\end{figure}

We take the first term in the two expansions (Eq.~(\ref{exp1}) and
  (\ref{exp2})) which is 
up to integrations 
(see Fig.~\ref{fig1})
\begin{eqnarray}
-\bar{B}(x)\; \left[ \Sigma(x,y) \;+\;\frac{1}{\tilde{G}} \delta(x-y)
\right] \;B(y)\;,
\end{eqnarray}
where $\Sigma(x,y)$ is the nucleon self-energy:
\begin{eqnarray}
\Sigma(x,y) &=& -N_c \;i\;\gamma^\mu \gamma^5\; {\tau_i} \; i
  \tilde{\Delta}_{0\; \mu \nu}^{i j}(x,y) \; i S_0(x,y)
  \;{\tau_j}\;\gamma^\nu \gamma^5\; \rm{sin}^2\theta \nonumber\\ &&-
  N_c \; i {\Delta_0} (x,y)\; i S_0(x,y)\; \rm{cos}^2\theta\;.
\end{eqnarray}
Here $N_c = 3$ is the number of colors (resulting from the trace over
color), and 
summation over repeated 
indices is to be
understood. The Fourier transform $\Sigma(p)$ of the self-energy is
decomposed according to
\begin{eqnarray}
\Sigma(p) &=& \Sigma_s(p^2) \;+\; \rlap/p \; \Sigma_v(p^2)\;.
\end{eqnarray}
The nucleon mass $M_B$ is then given by the vanishing of the inverse nucleon 
propagator:
\begin{eqnarray} 
\frac{1}{\tilde{G}} \;+\; \Sigma_s(M_B^2) \;+\; M_B \;\Sigma_v(M_B^2)
= 0\;.
\label{mass}
\end{eqnarray}
This condition generates dynamically the nucleon mass (one of the
predictions of the model) in terms of the theory parameters, and it is
similar in structure to the 
mass equations that 
determine the
masses in the meson sector~\cite{ER86,ERV94}.
Near the mass shell, the inverse nucleon propagator takes then the form:
\begin{eqnarray}
\left[ \Sigma(p) \;+\;
\frac{1}{\tilde{G}}\right] &\sim& (\rlap/p-M_B) \;Z^{-1}\;. 
\end{eqnarray}
Evidently, the nucleon has now acquired the desired status as a
dynamical degree of freedom in the problem. Here $Z$ is the
wave-function renormalization constant (see Sec.~\ref{renorm}) which prompts us to renormalize the nucleon field
according to $B = \sqrt{Z} \; B_{\rm ren}$.

\subsection{Regularization of divergent integrals and Ward identity}

Now we are in a place to discuss regularization. The self-energy and
the various Feynman diagrams in the problem involve the evaluation of
divergent integrals. Consequently, we are confronted with the question
of how to regularize these integrals. This issue emerged as a decisive
one in our analysis as we have attempted several regularization
schemes.  We started by adopting the method of four-momentum sharp
cut-off, but found it unsatisfactory as it yielded a violation of the
Ward-Takahashi identity. Guided by ``experience'', one can remove by
fiat the terms that violate gauge invariance, and thus conforms to
this identity~\cite{EJ98}. However, a more rigorous and solid method
is certainly desirable. Accordingly, we sought to regularize the
integrals using both the three-momentum sharp cut-off and the
Pauli-Villars methods.  The former is motivated by dispersion
theory~\cite{HK94} and leads, as we verified, to compliance with the
Ward identity.  Yet, we found that the most suitable regularization
scheme, in terms of rigor and convenience, is the Pauli-Villars
technique which we have established in this work as the standard method
for regularizing all divergent integrals.  This method consists of
introducing a fictitious propagator with some mass $M$ to cancel the
divergent contribution in the integral at large momentum values (see
the Appendix~\ref{appendix}).  As a matter of principle, the
Pauli-Villars mass which appears in the nucleon sector can be
different from the NJL
cut-off $\Lambda$ 
arising in the meson sector.
Nonetheless, to minimize the number of free
parameters, we elected to equate them,
$M = \Lambda$.
It is noteworthy here that all
observables (see Sec.~\ref{num} below) were found to be 
insensitive to the value of the
Pauli-Villars mass upholding the futility of using it as a free
parameter.

\begin{figure}
\includegraphics{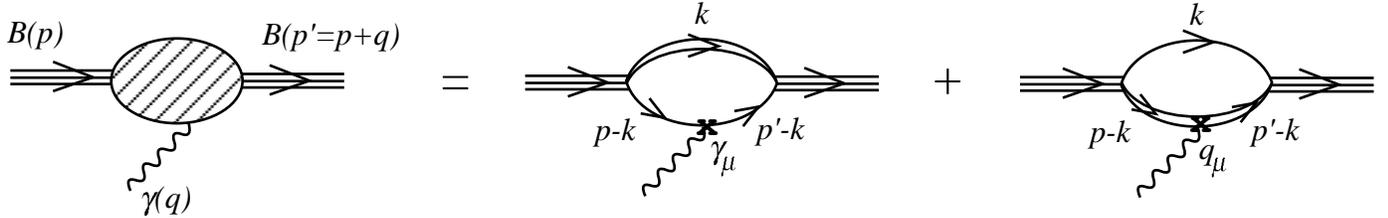}\\ 
\caption{\label{fig2} The Feynman diagrams for the electromagnetic
coupling which generate the nucleon electromagnetic
vertex.}
\end{figure}

In the process of testing gauge invariance (the Ward-Takahashi
identity), we have to determine the electromagnetic vertex of the
nucleon. This implies evaluating two kinds of diagrams depicted in
Fig.~\ref{fig2} where the nucleon can couple to the
electromagnetic field through either the quark or the diquark
propagators.  For the Ward-Takahashi identity to be satisfied, the
wave-function renormalization $Z$ must be equal to $Z_1$ where $Z_1$
is the electromagnetic vertex renormalization constant at $q^2 = 0$. Here,
$q = p^\prime - p$ (incoming photon) is the momentum
transfer. This condition is indeed satisfied for both the
Pauli-Villars and the three-momentum cut-off methods. By
calculating these diagrams at an arbitrary value of momentum transfer, we
derive the nucleon form factors from which we can extract the
electromagnetic radii and anomalous magnetic moments.

\subsection{Nucleon axial-vector vertex and the Goldberger-Treiman
relation}

\begin{figure}
\includegraphics{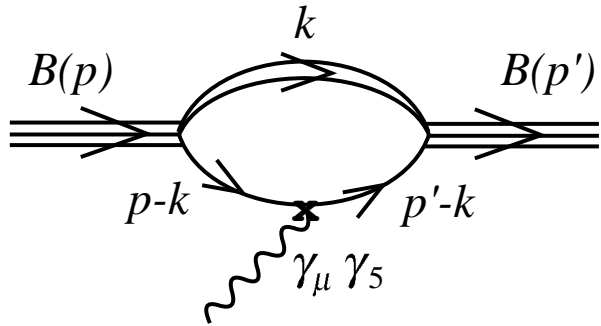}
\caption{\label{fig3} The Feynman diagram for the axial-vector
coupling which generates the quark contribution to the nucleon weak-interaction vertex.}
\end{figure}

We are in a position to calculate the weak-interaction
axial-vector vertex (Fig.~\ref{fig3}) to
determine the axial-vector form factors from which we can extract the
axial-vector coupling constant $g_A$. The $g_A$ is defined as the coefficient
of the $\gamma^\mu \gamma^5$ term of the axial-vector vertex at vanishing
momentum transfer. This vertex leads naturally to the Goldberger-Treiman
relation as can be seen by noticing that
\begin{eqnarray}
\vec{\mathcal A}^\pi_\mu (\Phi) = \frac {1}{F_\pi}\; \partial_\mu \vec{\Phi}\; +\;
{\cal O} (\Phi^3)\;.
\end{eqnarray}
Thus we obtain the following term for the pion-nucleon coupling:
\begin{eqnarray}
g_A \;\bar{B} \; \gamma^\mu \gamma^5 \frac{\vec{\tau}}{2}\cdot
\vec{\mathcal A}^\pi_\mu \;B \; \longrightarrow \; \frac {g_A}{F_\pi} \; \bar{B}\; \gamma^\mu \gamma^5
\frac{\vec{\tau}}{2}\cdot  \partial_\mu \vec{\Phi}\;B \;. 
\end{eqnarray}
But this term has to be identified with the pseudovector form of 
the Yukawa pion-nucleon coupling~\cite{SW86}: $ \frac{g_{\Phi N N}}{2M_B} \;\bar{B} \:\gamma^\mu \gamma^5
{\vec{\tau}}\cdot \partial_\mu \vec{\Phi} \;B$, which prompts us to
conclude that
\begin{eqnarray}
g_{\Phi N N}&=&\frac{M_B}{F_\pi}g_A\;.
\end{eqnarray}
This is nothing but the Goldberger-Treiman relation at the composite
hadron level. Note that this relation appears intact with no effort in
our treatment as opposed to large violations of up to 30\% in the
Bethe-Salpeter equation approach~\cite{HAOR97}. 

Finally, one must
mention that in addition to the weak-gauge-boson coupling through the
quark line (Fig.~\ref{fig3}), there is, only for the case of the axial-vector
diquark, a weak coupling through the diquark line. The
quantitative contribution of this diquark to the
Goldberger-Treiman relation remains to be investigated in a future work.

\section{Numerical study} 
\label{num}

Having thus far derived the structure of the problem, we
proceed to generate numerical results for our
model using the simpler case of only scalar diquarks, thereby
admitting the possibility of an intrinsic diquark form factor. Including
only scalar diquarks is not out of
place. Indeed, recent studies using scalar diquarks have reported
good results for most of the nucleon
observables~\cite{HW95,Keiner96a,HAOR97}. Moreover, there are
indications of scalar diquark dominance in the
nucleon~\cite{CRCP87,Rein90,Vogl90}. As a matter of fact, a very recent
 calculation using the Faddeev equation for three quark
states has concluded that axial-vector correlations, while
still important for magnetic properties, contribute at most no more than 10\%
to the structure of the nucleon~\cite{MBIY02}.

\subsection{Free parameters and basic quantities}
\label{param}

Tab.~\ref{tab1} provides the basic quantities in our model.
\setlength{\tabcolsep}{3mm}
\begin{table}
\caption{\label{tab1}Basic quantities in our model: The basic
quantities in the microscopic model are the constituent quark mass ($m_q$),
the scalar diquark mass ($M_S$), the cut-off $\Lambda$, and the quark-diquark coupling constant
$\tilde{G}$.}
\begin{ruledtabular}
\begin{tabular}{cccc}
$m_q$ & $M_S$ & $\Lambda$ & $\tilde{G}$\\
$.390$~GeV & $.600$~GeV & $.630$~GeV & $271.0$~GeV$^{-1}$
\end{tabular}
\end{ruledtabular}
\end{table}
As for free parameters, we have first the NJL coupling constant $G$
and the NJL cut-off $\Lambda$ which are fixed to yield the constituent
quark mass (quark condensate) through the NJL gap equation in the
meson sector~\cite{ER86,HK94,ERV94}. In this fashion, the $G$
constant decouples completely from the nucleon sector, the
sector of our interest.
The two diquark masses are also determined, in a consistent
manner, using the NJL model and the Bethe-Salpeter equation in the
diquark channels\cite{VW91,CRCP87}. In this context, the diquark
masses are simply poles, just as mesons, but in the quark-quark $T$
matrix. This leaves us with only one new free parameter in our
model: the quark-diquark coupling constant $\tilde{G}$. As can be
discerned, this model is well-constrained and yields a powerful
predictive strength.  
It has to 
be remarked here that in principle there is
another free parameter in the model
: $\theta$ as the mixing angle
for the two diquark contributions.
But this angle has no effect in the
present analysis as we consider only scalar diquarks
($\theta = 0$). Moreover, one must mention that while the quark and diquark
 masses and the cut-off $\Lambda$ are in principle fixed through the
 NJL model, small variations in their values are permissible as they still lead to
 consistent results within the NJL model. This adds a
 margin of freedom to these
 masses.

\subsection{Nucleon static properties}

Tab.~\ref{tab2} displays our predictions for some of the static
properties of the nucleon. Experimental values are taken from
Ref.~\cite{Dumbrajs83,particle00}. A mass of 0.94~GeV is obtained for
the nucleon through the mass equation Eq.~(\ref{mass}).  By fixing the
nucleon mass at this value, we would have eliminated the $\tilde{G}$
coupling constant from the problem and reached a theory with no more
free parameters. As for the binding energy of the nucleon, it is
estimated ($m_q = 0.390$~GeV and 
$M_S = 0.600$~GeV) as
$\Delta E_{\text{bin}} \equiv m_q + M_S - M_B = 50$~MeV, suggesting that
the nucleon is a loosely bound state of a quark and a diquark. This
prediction is consistent with other approaches of the NJL
model~\cite{HW95} or using the Faddeev equation~\cite{IBY93b}.

\begin{table}
\caption{\label{tab2}Model predictions: Some of the nucleon static
properties as predicted in the present calculation using the intrinsic
diquark form factor (IDFF) or without it. Experimental
values are taken from Ref.~\cite{Dumbrajs83,particle00}.}
\begin{ruledtabular}
\begin{tabular}{cccccccc}
& $\mu_p$ & $\mu_n$ & $g_A$ & $<r^2>^p_E$ & $<r^2>^n_E$
& $<r^2>^p_M$  & $<r^2>^n_M$\\
&&&&(fm$^2$)&(fm$^2$)&(fm$^2$)&(fm$^2$)\\
\hline 
Theory with IDFF& 1.57 & -.75 & .87 & .77& -.11& .82& .84\\
Theory without IDFF& 1.57 & -.75 & .87 & .68& -.19& .82& .85\\
Experiment&2.79 & -1.91 &  1.26 & .74  & -.12 & .74 & .77
\end{tabular}
\end{ruledtabular}
\end{table}

In the same table, we show the magnetic moments of the proton and the
neutron. Our treatment predicts a number that is two-third of the
experimental value for the proton and 
about
one-half of that for the
neutron. This is not a surprising result considering that we have not
included the axial-vector diquark in the present
calculation. Constituent quark models and other more sophisticated
approaches predict precisely that the axial-vector diquark inclusion
should add the missing one-third strength to the proton and the
missing one-half one to the neutron~\cite{HW95,Keiner96a}. So in this
context, this result is exactly what one should have expected from our
current analysis. For the same reason, the predicted value for the
axial-vector coupling $g_A$ of $0.87$ is significantly less than the
experimental one of $1.26$. While the scalar diquark cannot
couple to the weak interaction, the axial-vector one indeed does
couple to the weak gauge bosons adding strength to the
interaction. The magnetic moments and the axial-vector coupling
constant display a rather small sensitivity to the coupling constant
$\tilde{G}$. In fact the calculated values for the magnetic moments
are not that different from the predictions based on simple additive
models~\cite{HW95} suggesting a rather independence from the
details of the dynamics or the nucleon size.

Speaking of the nucleon size, it is nicely well-produced by our model:
the electric and magnetic radii for the proton and the neutron are
close to the experimental measurements. The negative charge radius of
the neutron has been suggested as an indication of a scalar diquark
clustering in the nucleon~\cite{DMW81}, and our treatment manifests
this conjecture in a dynamical model. These numbers point to a
physical picture of a ``heavy'' diquark at the center with a quark
rotating around it. By comparing the radii as calculated with and
without the intrinsic diquark form factor, we find that the
extended size of the diquark contributes a positive value of about
$0.10~\text{fm}^2$ for each of the proton and neutron electric
radii. Note that the scalar diquark has a positive charge and thus the
contribution is positive adding about $0.10~\text{fm}^2$ to the proton
radius and reducing the absolute value of the neutron one by
the same amount. We conclude here that the size of the diquark
contributes about 10\% of the proton radius and 40\% of the neutron
one. This confirms an earlier calculation using a static quark-exchange approximation~\cite{HW95}.  As
anticipated, the intrinsic form factor has virtually no effect on the
magnetic radii as the scalar diquark, as we shall see below, has a
negligible contribution to the magnetic form factors. Moreover, we have found the electromagnetic radii to be very sensitive to the
binding energy as well as, although implicitly through the
mass equation 
(Eq.(~\ref{mass})), the coupling constant
$\tilde{G}$. This indicates, as easily expected, the importance of the
details of the dynamics for the nucleon size. We should remark
here that the quantities
calculated here do not contain the pion contribution which becomes
significant for some physical quantities. It is estimated that for
$m_\pi = 138$~MeV, typical values for the pion corrections are of
order of 30\%\footnote{In the exact chiral limit ($m_\pi \rightarrow
0$), the pion contribution to the isovector charge radius diverges.}.

\subsection{Nucleon electric and magnetic form factors}

\begin{figure}[h]
\begin{center}
\includegraphics[totalheight=7.0in,angle=-90]{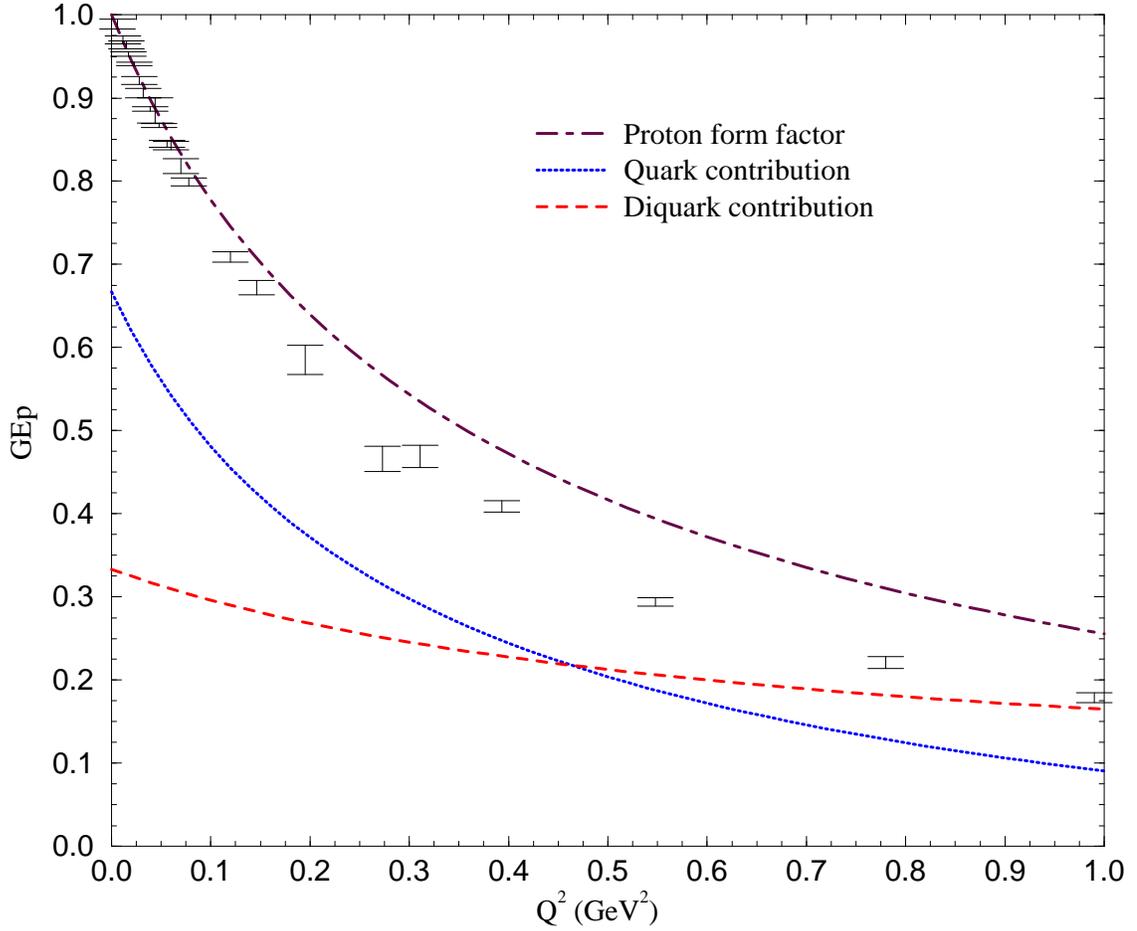}
\caption{\label{fig4} The proton electric form factor in comparison
with experimental data. The figure shows the quark contribution
(dotted line), the diquark contribution (dashed line) as well as
the full form factor (dotted-dashed line) as the sum of the two
contributions. Intrinsic structure of the scalar diquark is not
included here. Experimental
data from Ref.~\cite{Hohler76} are included.}
\end{center}
\end{figure}

\begin{figure}[h]
\begin{center}
\includegraphics[totalheight=7.0in,angle=-90]{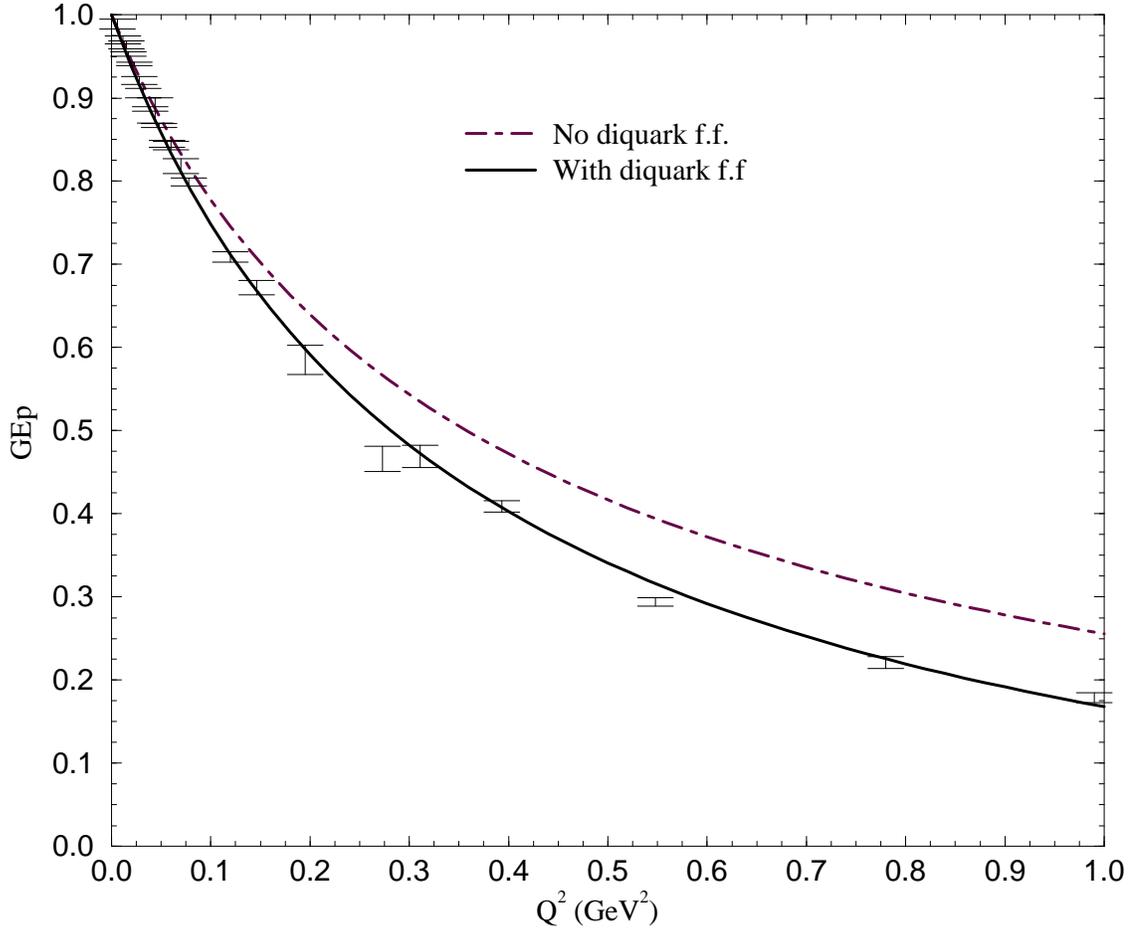}
\caption{\label{fig5} The proton electric form factor, with intrinsic
diquark form factor (diquark f.f.), in comparison with experimental data. The figure
shows the form factor with no intrinsic diquark form factor (dotted-dashed
line), and with the intrinsic form factor (solid line).
Experimental data from Ref.~\cite{Hohler76} are included.}
\end{center}
\end{figure}

Next we calculate the nucleon form factors. Fig.~\ref{fig4} displays
the proton electric form factor in comparison with experimental data
taken from Ref.~\cite{Hohler76}. The figure shows the quark
contribution, the diquark contribution in addition to the full form
factor (the sum of the two contributions), with no intrinsic diquark
form factor.
It is evident that our treatment reproduces beautifully the form
factor at low values of momentum transfer ($Q^2 \equiv - q^2$ where
$q$ is the momentum transfer). The discrepancy at higher values of
$Q^2$ begs for an understanding.
The figure suggests an explanation: the
diquark contribution
is almost constant implying a
rather localized diquark inside the proton. Although treated as an elementary
field in our theory, the diquark is a composite object and does have a
finite size. Our treatment needs to be adjusted to reflect the true
nature of the diquark by incorporating an intrinsic diquark form
factor. Fortunately, diquark form factors have been calculated
recently and in fact in the framework of the NJL
model~\cite{WBAR93}. Thus we readily add the intrinsic diquark form
factor to our treatment and produce the proton form factor
shown in Fig.~\ref{fig5}.  Impressively, the
calculations matches very well with the experimental data.

\begin{figure}[h]
\begin{center}
\includegraphics[totalheight=6.5in,angle=-90]{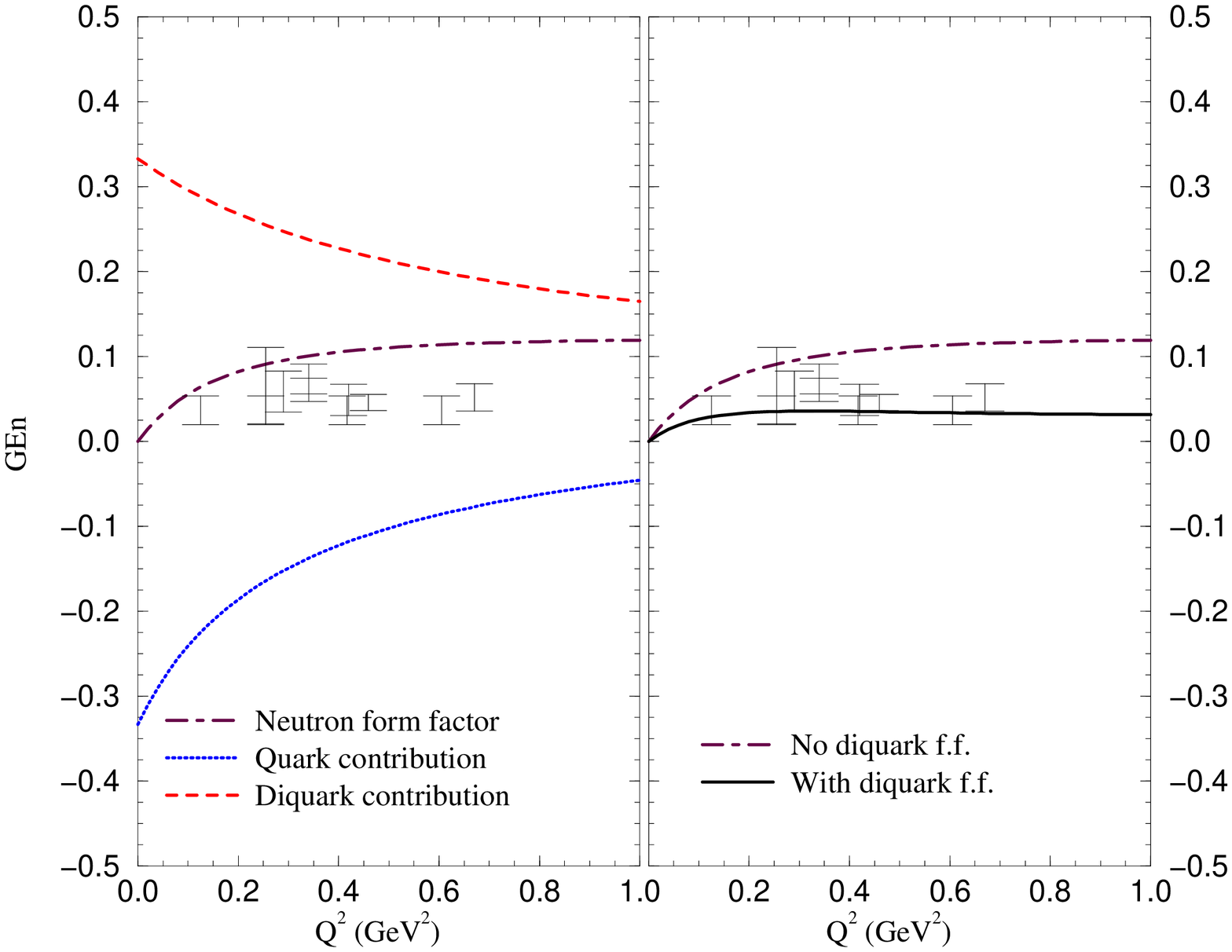}
\caption{\label{fig6} The neutron electric form factor in comparison
with experimental data. The left panel shows the quark contribution
(dotted line), the diquark contribution (dashed line) as well as
the full form factor (dotted-dashed line) as the sum of the two
contributions. The right panel displays the neutron form factor with
no intrinsic diquark form factor (dotted-dashed line), and with the
intrinsic form factor (solid line). Experimental data from
Ref.~\cite{Eden94,Bruins95,Ostrick99,Rohe99,Zhu01} are included.}
\end{center}
\end{figure}

The neutron electric form factor tells a similar story. In the left
panel of Fig.~\ref{fig6}, just as in Fig.~\ref{fig4}, we display the
neutron
form factor with its quark and diquark contributions. Clearly, the
quark contribution is negative in value (d-quark) and thus cancels
much of the diquark contribution leading to a small form factor. There
is once more a discrepancy compared to experimental data that is
largely eliminated once we include the intrinsic form factor as
exhibited in the right panel of the same figure. It is noteworthy
here that the neutron form factor is a potent test of any treatment as
it is a delicate cancellation of two large
contributions~\cite{Keiner96a,Keiner96b}. Saliently, the cancellation
is naturally produced in our study. The experimental data are obtained
from Ref.~\cite{Eden94,Bruins95,Ostrick99,Rohe99,Zhu01}.

\begin{figure}[h]
\begin{center}
\includegraphics[totalheight=7.0in,angle=-90]{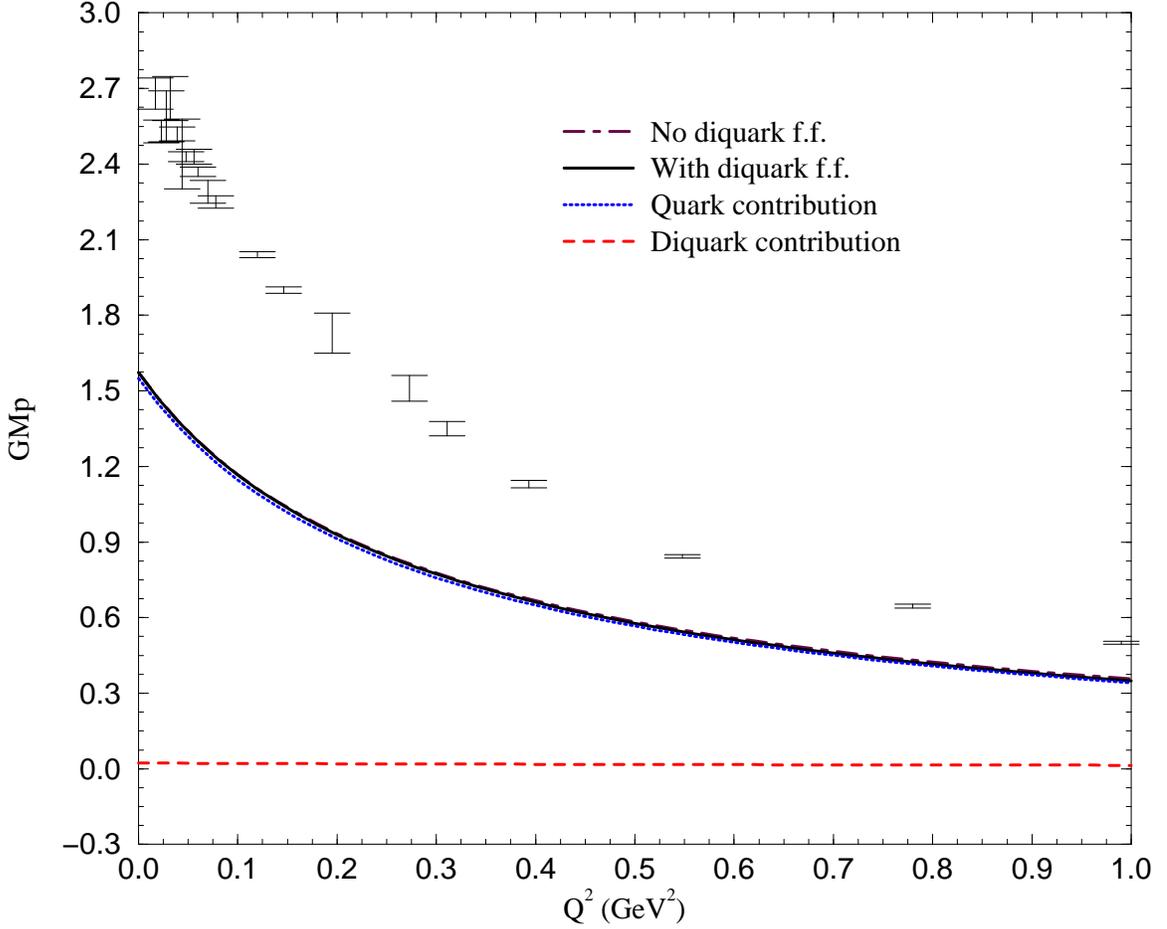}
\caption{\label{fig7} The proton magnetic form factor in comparison
with experimental data. The figure shows the proton magnetic form
factor as calculated with (solid line) or without (dotted-dashed line)
the intrinsic diquark form factor. The figure also includes the quark
(dotted line) and the diquark (dashed line) contributions to this
form factor. Note that the three curves (apart form the diquark
contribution) are very similar. Experimental data from Ref.~\cite{Hohler76} are
included.}
\end{center}
\end{figure}

In Fig.~\ref{fig7}, we present the proton magnetic form factor as
calculated with or without the intrinsic diquark form factor. The
figure also contains the quark and diquark contributions. Unmistakably, the scalar diquark contribution is
virtually vanishing due to the lack of an intrinsic spin.
Nevertheless, there is a very small contribution due to a small
orbital angular-momentum effect in the bound quark-diquark system.
Since the diquark contribution is negligible, the inclusion of the
intrinsic diquark form factor does not alter our prediction and the
form factor is determined to be almost purely from a quark
origin. This suggests the need for the axial-vector
diquark, which does have an intrinsic spin, to supplement the quark
contribution and to provide the missing one-third strength compared to
the experimental data~\cite{Hohler76}. The figure also indicates a
convergence of our calculation, mainly from a quark origin,
and the experimental data at large values of $Q^2$. Such result
suggests that this form factor is almost purely from a quark origin in
this regime. This is anticipated due to the finite size of the
axial-vector diquark which probably can have a significant
contribution but only for smaller values of $Q^2$.

\begin{figure}[h]
\begin{center}
\includegraphics[totalheight=7.0in,angle=-90]{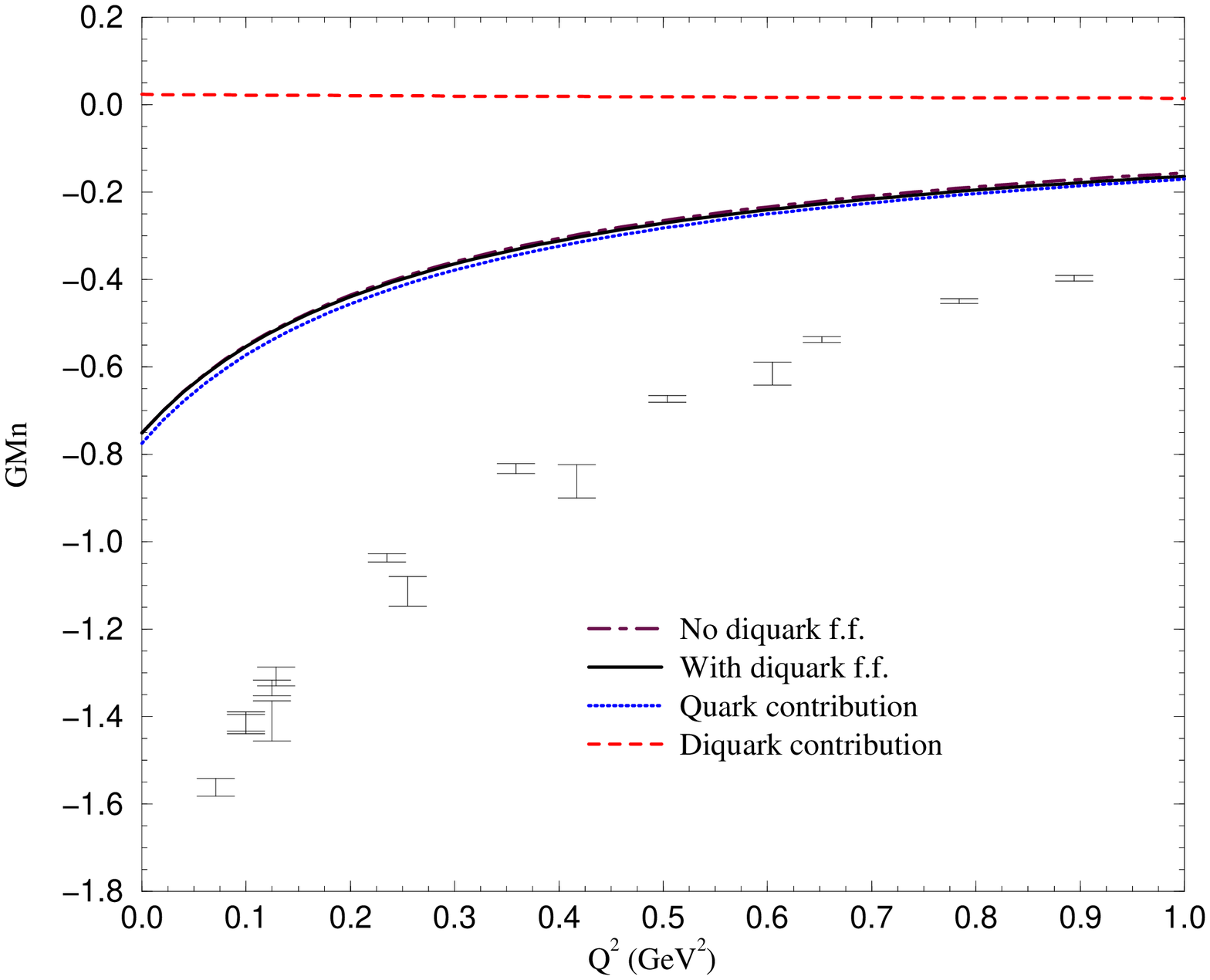}
\caption{\label{fig8} The neutron magnetic form factor in comparison
with experimental data. The figure shows the neutron magnetic form
factor as calculated with (solid line) or without (dotted-dashed line)
the intrinsic diquark form factor. The figure also includes the quark
(dotted line) and the diquark (dashed line) contributions to this
form factor. Note that the three curves (apart form the diquark
contribution) are very similar. Experimental data from
Ref.~\cite{Anklin94,Bruins95,Anklin98,Golak00,Xu00,Kubon01} are
included.}
\end{center}
\end{figure}

The neutron magnetic form factor describes a similar narrative to that
of the proton but here the missing strength (one-half) is larger as
can be seen in Fig.~\ref{fig8}.  As stated earlier, these specific missing
strengths are predicted due to the absence of the
axial-vector diquark in the present analysis. The experimental data in
Fig.~\ref{fig8} are obtained from
Ref.~\cite{Anklin94,Bruins95,Anklin98,Golak00,Xu00,Kubon01}.

\section{Conclusions}
\label{concl}

In this paper we tackled the nucleon structure and the challenging
problem of understanding the origin and nature of the nuclear force by
deriving a meson-nucleon Lagrangian using the path-integral method of
hadronization.  We started from a microscopic model of quarks and
diquarks where the gluonic degrees of freedom have been integrated
out. The nucleon was conceived as quark-diquark
correlations and only two kinds of diquarks were found relevant for
its structure. These are the scalar isoscalar and the axial-vector isovector
diquarks.  Composite meson and nucleon fields were introduced by the
methods of path-integral bosonization and fermionization to rewrite
the problem in terms of the physical meson and nucleon degrees of
freedom.  This yielded an effective chiral meson-nucleon Lagrangian
after using a loop and derivative expansions of the resulting
quark/diquark determinants. The divergent loop diagrams were
regularized using gauge-invariant regularization schemes and the
Ward-Takahashi identity and the Goldberger-Treiman relation were
verified.

An extensive set of nucleon observables were calculated for the first time
on the basis of the path-integral hadronization approach. 
Indeed, many of the nucleon
physical properties such as mass, coupling constants, electromagnetic
radii, anomalous magnetic moments, and form factors have been
determined from a model of essentially one free parameter. By taking
into account the intrinsic diquark form factor, we established
a remarkable agreement with the experimental data for the nucleon size
and the electric form factors, while our calculations show missing
strengths for the magnetic form factors and the axial-vector coupling
constant. The discrepancy is likely due to the absence of the
axial-vector diquark in the present numerical study.

This work is part of an ambitious program of using path-integral
techniques and QCD-based effective field theories to study baryon
structure and to derive a full-fledged nuclear force. The final goal
of the program is a derivation of an effective field theory for the
nuclear force, namely, quantum hydrodynamics (QHD) from quark
dynamics.  As for the future, we plan to attain a numerical study
using both the scalar and axial-vector diquarks including their
intrinsic form factors . This is a challenging and rather difficult
task due to the axial-vector diquark intricate structure as a particle
of one unit spin and isospin. Some of the ensuing complications are
the axial-vector diquark direct coupling to the weak interaction and
the electroweak scalar-axial-vector transitions.  At a later stage, we
plan to generalize our approach to chiral SU(3) symmetry to study the
structure of the baryon octet.

\appendix*
\section{Analytical expressions using the Pauli-Villars regularization method\label{appendix}}

We include here analytical expressions for some of the principal
formulae in our treatment. 

\subsection{Self-energy and wave-function renormalization\label{renorm}}

The self-energy is depicted by the Feynman diagram of Fig.~\ref{fig1}
and is given by the expression:

\begin{eqnarray}
\Sigma(p) &=& N_c\; (- i) \int \frac{d^4k}{(2\pi)^4}\;
\frac{i\L(\rlap/p - \rlap/k + m_q\R)}{(p - k)^2 - m_q^2} \; \L[
\frac{i}{k^2 - M_S^2} - \frac{i}{k^2 - \Lambda^2}\R]\;,\nonumber\\
\nonumber\\ 
&=& N_c \; \frac{1}{(4\pi)^2} \int^1_0 dx \;\L[\rlap/p (1-x) + m_q\R]\;
\ln \L[\frac{\Delta(\Lambda,q^2 = 0)}{\Delta(M_S,q^2 = 0)}\R]\;.
\end{eqnarray}
Here $p^\mu$ is the momentum of the nucleon 
taken to be on the mass shell,
 $N_c = 3$ is the number
of colors, and
\begin{eqnarray}
\Delta(M,q^2) \equiv m_q^2 x + M^2 (1-x) - p^2 x (1-x) + q^2 x^2 y
(y-1)\;.
\end{eqnarray}
Note that we have used the Pauli Villars method for regularizing the
divergent integral by incorporating the propagator $\frac{i}{k^2 -
\Lambda^2}$ of a fictitious
scalar particle with mass $\Lambda$ in the above expression.

The wave-function renormalization $Z$ is obtained through the
derivative $\frac{\partial \Sigma(p)}{\partial p_\mu}
{\bigg|}_{\rlap/p \rightarrow M_B}$, leading to the expression:
\begin{eqnarray}
Z^{-1} &=& N_c \;\frac{1}{(4\pi)^2} \int^1_0 dx \; \bigg\{ \;(1-x) \ln
\L[\frac{\Delta(\Lambda,q^2=0)}{\Delta(M_S,q^2=0)}\R] - \nonumber\\
&&2 x (1-x)\:M_B \:\L[ M_B (1-x) + m_q\R]\; \L[
\frac{1}{\Delta(\Lambda,q^2=0)} - \frac{1}{\Delta(M_S,q^2=0)}
\R]\;\bigg\}\;.
\end{eqnarray}

\subsection{Electromagnetic interaction}

The photon couples to both the quark and the diquark lines leading to
two contributions to the nucleon electromagnetic vertex.

\subsubsection{Quark contribution}

This is the contribution represented by the left part of
Fig.~\ref{fig2} and is given by:
\begin{eqnarray}
-\Lambda^\mu_{\gamma q} &=& -\;N_c\; Q_q\; \int \frac{d^4k}{(2\pi)^4}\;
\L[ \frac{i}{k^2 - M_S^2} - \frac{i}{k^2 - \Lambda^2}\R]
\frac{i\L(\rlap/p^\prime - \rlap/k + m_q\R)}{(p^\prime - k)^2 - m_q^2}
\;\gamma^\mu\; \frac{i\L(\rlap/p - \rlap/k + m_q\R)}{(p - k)^2 -
m_q^2}\;, \nonumber\\ 
\nonumber\\ 
&=& F_1^q(q^2) \; \gamma^\mu + F_2^q(q^2) \;
\frac{i \sigma^{\mu\nu} \:q_\nu}{2 M_B}\;,
\end{eqnarray}
where $F_1^q (q^2)$ and $F_2^q(q^2)$ are the quark contributions to
the nucleon form factor and are given by:
\begin{eqnarray}
 F_1^q(q^2) &=& Z \;N_c \; Q_q\; \frac{1}{(4\pi)^2} \; \int^1_0 dx
\int^1_0 dy \; \; \bigg\{ \; (1-x) \ln
\L[\frac{\Delta(\Lambda,q)}{\Delta(M_S,q)}\R] - \nonumber\\ &&\bigg[ 2
x (1-x)M_B \:\L[ M_B (1-x) + m_q\R] + q^2 x^2 (1-y)\bigg] \L[
\frac{1}{\Delta(\Lambda,q)} - \frac{1}{\Delta(M_S,q)} \R]\;
\bigg\}\;,\nonumber\\ \\ 
F_2^q(q^2) &=& Z \;N_c \; Q_q\;
\frac{1}{(4\pi)^2} \; \int^1_0 dx \int^1_0 dy \;\; \nonumber \\
\; && \hspace{2.5cm} \; \times \; 2 x^2 M_B \: \L[ M_B (1-x) + m_q\R] \; \L[
\frac{1}{\Delta(\Lambda,q)} - \frac{1}{\Delta(M_S,q)} \R]\; \;.
\end{eqnarray}
In the above expressions, $p^{\mu}$ ($p^{\prime \mu}$) is the momentum
of the incoming (outgoing) nucleon, $q^\mu = p^{\prime \mu} - p^{\mu}$
is the momentum transfer, and $Q_q = \text{diag}\L(2/3,-1/3\R)$
is the quark charge. Note that we have
included the wave-function renormalization constant $Z$ in the above
expressions.

\subsubsection{Scalar diquark contribution}

This is the contribution represented by the right part of
Fig.~\ref{fig2} and is given by:
\begin{eqnarray}
-\Lambda^\mu_{\gamma D} &=& -\;N_c\; Q_S\; \int
\frac{d^4k}{(2\pi)^4}\;\bigg\{ \frac{i}{(p - k)^2 - M_S^2} \;
\frac{i\L(\rlap/k + m_q\R)}{k^2 - m_q^2}\; \frac{i}{(p^\prime - k)^2 -
M_S^2} -\nonumber\\ &&\hspace{2.0cm} \frac{i}{(p - k)^2 - \Lambda^2}
\; \frac{i\L(\rlap/k + m_q\R)}{k^2 - m_q^2}\; \frac{i}{(p^\prime -
k)^2 - \Lambda^2} \;\bigg\}\; \L(p^\mu - p^{\prime\;\mu} - 2
k^\mu\R)\;,\nonumber\\ 
\nonumber\\
&=& F_1^D(q^2) \; \gamma^\mu + F_2^D(q^2) \;
\frac{i \sigma^{\mu\nu} \:q_\nu}{2 M_B}\;,
\end{eqnarray}
where
\begin{eqnarray}
 F_1^D(q^2) &=& Z N_c \; Q_S\; \frac{1}{(4\pi)^2} \; \int^1_0 dx
\int^1_0 dy \; \; \bigg\{ \; (1-x) \ln
\L[\frac{\Delta(\Lambda,q)}{\Delta(M_S,q)}\R] - \nonumber\\ &&\bigg[ 2
x (1-x)M_B \: \L[ M_B (1-x) + m_q\R] \bigg] \L[
\frac{1}{\Delta(\Lambda,q)} - \frac{1}{\Delta(M_S,q)} \R]\; \bigg\}
\;,\\ \nonumber\\ F_2^D(q^2) &=& Z N_c \; Q_S\; \frac{1}{(4\pi)^2} \;
\int^1_0 dx \int^1_0 dy \;\;\bigg\{ \nonumber \\ \; && \hspace{1cm} \;
2 x (1-x) M_B \: \L[ M_B (1-x) + m_q\R] \; \L[
\frac{1}{\Delta(\Lambda,q)} - \frac{1}{\Delta(M_S,q)} \R]\; \bigg\}\;.
\end{eqnarray}
Here $Q_S = 1/3$ is the diquark charge.

\subsubsection{Form factor and Ward-Takahashi identity}

The full nucleon electromagnetic form factor is the sum of the quark
and diquark contributions. Note that at $q^2 = 0$, the sum of these
two pieces conforms to the Ward-Takahashi identity as it yields $Z\: \L(Q_q +
Q_S\R) \:\frac{1}{Z} = \text{diag}\L(1,0\R) = Q_N$ with the correct normalization for the
nucleon electric charge.

\subsection{Axial-vector coupling constant $g_A$}

The axial-vector coupling constant $g_A$ (with only scalar diquarks) is determined from the
axial-vector vertex as represented by Fig~\ref{fig3}. Note that
since the scalar diquark cannot couple to the weak interaction, there
is no contribution to $g_A$ from a direct coupling to the diquark
line. The full axial-vector vertex is given by:
\begin{eqnarray}
{\mathcal A}^{\mu}_{\text{axial}}& =& -\;N_c\; \int \frac{d^4k}{(2\pi)^4}\; i \L[
\frac{i}{k^2 - M_S^2} - \frac{i}{k^2 - \Lambda^2}\R]
\frac{i\L(\rlap/p^\prime - \rlap/k + m_q\R)}{(p^\prime - k)^2 - m_q^2}
\;\gamma^\mu \gamma^5\; \frac{i\L(\rlap/p - \rlap/k + m_q\R)}{(p -
k)^2 - m_q^2}\;.\nonumber\\
\end{eqnarray}
The $g_A$ is defined as the coefficient
of the $\gamma^\mu \gamma^5$ term of the axial-vector vertex at $q^2 =
0$. This yields after evaluating this vertex
\begin{eqnarray}
g_A &=& - Z\; N_c \; \frac{1}{(4\pi)^2} \; \int^1_0 dx \int^1_0 dy \;
\; \bigg\{ \; x \ln
\L[\frac{\Delta(\Lambda,q^2=0)}{\Delta(M_S,q^2=0)}\R] - \nonumber\\
\nonumber\\ && \hspace{2.1cm} x\:\L[ M_B (1-x) + m_q\R]^2 \L[
\frac{1}{\Delta(\Lambda,q^2=0)} - \frac{1}{\Delta(M_S,q^2=0)} \R]\;
\bigg\}\;.
\end{eqnarray}
\begin{acknowledgments}
L.J.A. acknowledges the support of a joint
fellowship from the Japan Society for the Promotion of Science and the
United States National Science Foundation. D.E. and A.H. thanks W. Bentz for fruitful discussions on the role
of axial-vector diquarks. A.H. also thanks A.W. Thomas for discussions on chiral corrections 
and hospitality during his stay at CSSM Adelaide. 
\end{acknowledgments}
\bibliography{nucleon}

\begin{thebibliography}{72}
\expandafter\ifx\csname natexlab\endcsname\relax\def\natexlab#1{#1}\fi
\expandafter\ifx\csname bibnamefont\endcsname\relax
  \def\bibnamefont#1{#1}\fi
\expandafter\ifx\csname bibfnamefont\endcsname\relax
  \def\bibfnamefont#1{#1}\fi
\expandafter\ifx\csname citenamefont\endcsname\relax
  \def\citenamefont#1{#1}\fi
\expandafter\ifx\csname url\endcsname\relax
  \def\url#1{\texttt{#1}}\fi
\expandafter\ifx\csname urlprefix\endcsname\relax\def\urlprefix{URL }\fi
\providecommand{\bibinfo}[2]{#2}
\providecommand{\eprint}[2][]{\url{#2}}

\bibitem[{\citenamefont{Weinberg}(1967)}]{Wein67}
\bibinfo{author}{\bibfnamefont{S.}~\bibnamefont{Weinberg}},
  \bibinfo{journal}{Phys.\ Rev.\ Lett.} \textbf{\bibinfo{volume}{18}},
  \bibinfo{pages}{188} (\bibinfo{year}{1967}).

\bibitem[{\citenamefont{S.~Coleman and Zumino}(1969)}]{CWZ69}
\bibinfo{author}{\bibfnamefont{J.~W.} \bibnamefont{S.~Coleman}}
  \bibnamefont{and} \bibinfo{author}{\bibfnamefont{B.}~\bibnamefont{Zumino}},
  \bibinfo{journal}{Phys.\ Rev.} \textbf{\bibinfo{volume}{177}},
  \bibinfo{pages}{2239} (\bibinfo{year}{1969}).

\bibitem[{\citenamefont{C.~Callan and Zumino}(1969)}]{CCWZ69}
\bibinfo{author}{\bibfnamefont{J.~W.} \bibnamefont{C.~Callan},
  \bibfnamefont{S.~Coleman}} \bibnamefont{and}
  \bibinfo{author}{\bibfnamefont{B.}~\bibnamefont{Zumino}},
  \bibinfo{journal}{Phys.\ Rev.} \textbf{\bibinfo{volume}{177}},
  \bibinfo{pages}{2247} (\bibinfo{year}{1969}).

\bibitem[{\citenamefont{Ebert and Volkov}(1981)}]{EV81}
\bibinfo{author}{\bibfnamefont{D.}~\bibnamefont{Ebert}} \bibnamefont{and}
  \bibinfo{author}{\bibfnamefont{M.}~\bibnamefont{Volkov}},
  \bibinfo{journal}{Fortschr.\ Phys.} \textbf{\bibinfo{volume}{29}},
  \bibinfo{pages}{35} (\bibinfo{year}{1981}).

\bibitem[{\citenamefont{Nambu and Jona-Lasinio}(1961)}]{NJL1}
\bibinfo{author}{\bibfnamefont{Y.}~\bibnamefont{Nambu}} \bibnamefont{and}
  \bibinfo{author}{\bibfnamefont{G.}~\bibnamefont{Jona-Lasinio}},
  \bibinfo{journal}{Phys.\ Rev.} \textbf{\bibinfo{volume}{122}},
  \bibinfo{pages}{345} (\bibinfo{year}{1961}), \bibinfo{note}{ibid. 124,246
  (1961)}.

\bibitem[{\citenamefont{Ebert and Reinhardt}(1986)}]{ER86}
\bibinfo{author}{\bibfnamefont{D.}~\bibnamefont{Ebert}} \bibnamefont{and}
  \bibinfo{author}{\bibfnamefont{H.}~\bibnamefont{Reinhardt}},
  \bibinfo{journal}{Nucl. Phys.} \textbf{\bibinfo{volume}{B271}},
  \bibinfo{pages}{188} (\bibinfo{year}{1986}), \bibinfo{note}{and references
  therein}.

\bibitem[{\citenamefont{Ida and Kobayashi}(1966)}]{IK66}
\bibinfo{author}{\bibfnamefont{M.}~\bibnamefont{Ida}} \bibnamefont{and}
  \bibinfo{author}{\bibfnamefont{R.}~\bibnamefont{Kobayashi}},
  \bibinfo{journal}{Prog. Theor. Phys.} \textbf{\bibinfo{volume}{36}},
  \bibinfo{pages}{846} (\bibinfo{year}{1966}).

\bibitem[{\citenamefont{Lichtenberg et~al.}(1968)}]{Lich68}
\bibinfo{author}{\bibfnamefont{D.}~\bibnamefont{Lichtenberg}}
  \bibnamefont{et~al.}, \bibinfo{journal}{Phys. Rev.}
  \textbf{\bibinfo{volume}{167}}, \bibinfo{pages}{1535} (\bibinfo{year}{1968}).

\bibitem[{\citenamefont{Vogl and Weise}(1991)}]{VW91}
\bibinfo{author}{\bibfnamefont{U.}~\bibnamefont{Vogl}} \bibnamefont{and}
  \bibinfo{author}{\bibfnamefont{W.}~\bibnamefont{Weise}},
  \bibinfo{journal}{Prog. Part. Nucl. Phys.} \textbf{\bibinfo{volume}{27}},
  \bibinfo{pages}{195} (\bibinfo{year}{1991}).

\bibitem[{\citenamefont{Anselmino et~al.}(1993)\citenamefont{Anselmino,
  Predazzi, Ekelin, Fredriksson, and Lichtenberg}}]{APEFL93}
\bibinfo{author}{\bibfnamefont{M.}~\bibnamefont{Anselmino}},
  \bibinfo{author}{\bibfnamefont{E.}~\bibnamefont{Predazzi}},
  \bibinfo{author}{\bibfnamefont{S.}~\bibnamefont{Ekelin}},
  \bibinfo{author}{\bibfnamefont{S.}~\bibnamefont{Fredriksson}},
  \bibnamefont{and} \bibinfo{author}{\bibfnamefont{D.~B.}
  \bibnamefont{Lichtenberg}}, \bibinfo{journal}{Rev. Mod. Phys.}
  \textbf{\bibinfo{volume}{65}}, \bibinfo{pages}{1199} (\bibinfo{year}{1993}).

\bibitem[{\citenamefont{Anselmino and Predazzi}(1989)}]{Ansel89}
\bibinfo{editor}{\bibfnamefont{M.}~\bibnamefont{Anselmino}} \bibnamefont{and}
  \bibinfo{editor}{\bibfnamefont{E.}~\bibnamefont{Predazzi}}, eds.,
  \emph{\bibinfo{title}{Diquarks. Proceedings, Workshop, Turin, Italy, October
  24- 26, 1988}} (\bibinfo{publisher}{World Scientific},
  \bibinfo{address}{Singapore}, \bibinfo{year}{1989}).

\bibitem[{\citenamefont{Donnachie and Landshoff}(1980)}]{DonL80}
\bibinfo{author}{\bibfnamefont{A.}~\bibnamefont{Donnachie}} \bibnamefont{and}
  \bibinfo{author}{\bibfnamefont{P.~V.} \bibnamefont{Landshoff}},
  \bibinfo{journal}{Phys. Lett.} \textbf{\bibinfo{volume}{B95}},
  \bibinfo{pages}{437} (\bibinfo{year}{1980}).

\bibitem[{\citenamefont{Close and Roberts}(1981)}]{Close81}
\bibinfo{author}{\bibfnamefont{F.~E.} \bibnamefont{Close}} \bibnamefont{and}
  \bibinfo{author}{\bibfnamefont{R.~G.} \bibnamefont{Roberts}},
  \bibinfo{journal}{Z. Phys.} \textbf{\bibinfo{volume}{C8}},
  \bibinfo{pages}{57} (\bibinfo{year}{1981}).

\bibitem[{\citenamefont{Fredriksson et~al.}(1982)\citenamefont{Fredriksson,
  Jandel, and Larsson}}]{FSJMLT82}
\bibinfo{author}{\bibfnamefont{S.}~\bibnamefont{Fredriksson}},
  \bibinfo{author}{\bibfnamefont{M.}~\bibnamefont{Jandel}}, \bibnamefont{and}
  \bibinfo{author}{\bibfnamefont{T.}~\bibnamefont{Larsson}},
  \bibinfo{journal}{Z. Phys.} \textbf{\bibinfo{volume}{C14}},
  \bibinfo{pages}{35} (\bibinfo{year}{1982}).

\bibitem[{\citenamefont{Kroll et~al.}(1991)\citenamefont{Kroll, Schurmann, and
  Schweiger}}]{KSS91}
\bibinfo{author}{\bibfnamefont{P.}~\bibnamefont{Kroll}},
  \bibinfo{author}{\bibfnamefont{M.}~\bibnamefont{Schurmann}},
  \bibnamefont{and}
  \bibinfo{author}{\bibfnamefont{W.}~\bibnamefont{Schweiger}},
  \bibinfo{journal}{Z. Phys.} \textbf{\bibinfo{volume}{A338}},
  \bibinfo{pages}{339} (\bibinfo{year}{1991}).

\bibitem[{\citenamefont{Stech}(1987)}]{Stech87}
\bibinfo{author}{\bibfnamefont{B.}~\bibnamefont{Stech}},
  \bibinfo{journal}{Phys. Rev.} \textbf{\bibinfo{volume}{D36}},
  \bibinfo{pages}{975} (\bibinfo{year}{1987}).

\bibitem[{\citenamefont{Neubert and Stech}(1991)}]{Neu91}
\bibinfo{author}{\bibfnamefont{M.}~\bibnamefont{Neubert}} \bibnamefont{and}
  \bibinfo{author}{\bibfnamefont{B.}~\bibnamefont{Stech}},
  \bibinfo{journal}{Phys. Rev.} \textbf{\bibinfo{volume}{D44}},
  \bibinfo{pages}{775} (\bibinfo{year}{1991}).

\bibitem[{\citenamefont{Faddeev}(1961{\natexlab{a}})}]{Fad61a}
\bibinfo{author}{\bibfnamefont{L.}~\bibnamefont{Faddeev}},
  \bibinfo{journal}{Zh. Eksp. Theor. Fiz} \textbf{\bibinfo{volume}{39}}
  (\bibinfo{year}{1961}{\natexlab{a}}).

\bibitem[{\citenamefont{Faddeev}(1961{\natexlab{b}})}]{Fad61b}
\bibinfo{author}{\bibfnamefont{L.}~\bibnamefont{Faddeev}},
  \bibinfo{journal}{Sov. Phys. JEPT} \textbf{\bibinfo{volume}{12}},
  \bibinfo{pages}{1014} (\bibinfo{year}{1961}{\natexlab{b}}).

\bibitem[{\citenamefont{Alkofer et~al.}(1992)\citenamefont{Alkofer, Reinhardt,
  Weigel, and Zuckert}}]{ARWZ92}
\bibinfo{author}{\bibfnamefont{R.}~\bibnamefont{Alkofer}},
  \bibinfo{author}{\bibfnamefont{H.}~\bibnamefont{Reinhardt}},
  \bibinfo{author}{\bibfnamefont{H.}~\bibnamefont{Weigel}}, \bibnamefont{and}
  \bibinfo{author}{\bibfnamefont{U.}~\bibnamefont{Zuckert}},
  \bibinfo{journal}{Phys. Rev. Lett.} \textbf{\bibinfo{volume}{69}},
  \bibinfo{pages}{1874} (\bibinfo{year}{1992}).

\bibitem[{\citenamefont{Ishii et~al.}(1993{\natexlab{a}})\citenamefont{Ishii,
  Bentz, and Yazaki}}]{IBY93a}
\bibinfo{author}{\bibfnamefont{N.}~\bibnamefont{Ishii}},
  \bibinfo{author}{\bibfnamefont{W.}~\bibnamefont{Bentz}}, \bibnamefont{and}
  \bibinfo{author}{\bibfnamefont{K.}~\bibnamefont{Yazaki}},
  \bibinfo{journal}{Phys. Lett.} \textbf{\bibinfo{volume}{B318}},
  \bibinfo{pages}{26} (\bibinfo{year}{1993}{\natexlab{a}}).

\bibitem[{\citenamefont{Ishii et~al.}(1993{\natexlab{b}})\citenamefont{Ishii,
  Bentz, and Yazaki}}]{IBY93b}
\bibinfo{author}{\bibfnamefont{N.}~\bibnamefont{Ishii}},
  \bibinfo{author}{\bibfnamefont{W.}~\bibnamefont{Bentz}}, \bibnamefont{and}
  \bibinfo{author}{\bibfnamefont{K.}~\bibnamefont{Yazaki}},
  \bibinfo{journal}{Phys. Lett.} \textbf{\bibinfo{volume}{B301}},
  \bibinfo{pages}{165} (\bibinfo{year}{1993}{\natexlab{b}}).

\bibitem[{\citenamefont{Meyer}(1994)}]{Meyer94}
\bibinfo{author}{\bibfnamefont{H.}~\bibnamefont{Meyer}},
  \bibinfo{journal}{Phys. Lett.} \textbf{\bibinfo{volume}{B337}},
  \bibinfo{pages}{37} (\bibinfo{year}{1994}),
  \eprint[http://arXiv.org/abs]{nucl-th/9407003}.

\bibitem[{\citenamefont{Hellstern and Weiss}(1995)}]{HW95}
\bibinfo{author}{\bibfnamefont{G.}~\bibnamefont{Hellstern}} \bibnamefont{and}
  \bibinfo{author}{\bibfnamefont{C.}~\bibnamefont{Weiss}},
  \bibinfo{journal}{Phys. Lett.} \textbf{\bibinfo{volume}{B351}},
  \bibinfo{pages}{64} (\bibinfo{year}{1995}),
  \eprint[http://arXiv.org/abs]{hep-ph/9502217}.

\bibitem[{\citenamefont{Salpeter}(1952)}]{Salpeter52}
\bibinfo{author}{\bibfnamefont{E.~E.} \bibnamefont{Salpeter}},
  \bibinfo{journal}{Phys. Rev.} \textbf{\bibinfo{volume}{87}},
  \bibinfo{pages}{328} (\bibinfo{year}{1952}).

\bibitem[{\citenamefont{Keiner}(1996{\natexlab{a}})}]{Keiner96a}
\bibinfo{author}{\bibfnamefont{V.}~\bibnamefont{Keiner}}, \bibinfo{journal}{Z.
  Phys.} \textbf{\bibinfo{volume}{A354}}, \bibinfo{pages}{87}
  (\bibinfo{year}{1996}{\natexlab{a}}),
  \eprint[http://arXiv.org/abs]{hep-ph/9509284}.

\bibitem[{\citenamefont{Keiner}(1996{\natexlab{b}})}]{Keiner96b}
\bibinfo{author}{\bibfnamefont{V.}~\bibnamefont{Keiner}},
  \bibinfo{journal}{Phys. Rev.} \textbf{\bibinfo{volume}{C54}},
  \bibinfo{pages}{3232} (\bibinfo{year}{1996}{\natexlab{b}}),
  \eprint[http://arXiv.org/abs]{hep-ph/9603226}.

\bibitem[{\citenamefont{Salpeter and Bethe}(1951)}]{SB51}
\bibinfo{author}{\bibfnamefont{E.~E.} \bibnamefont{Salpeter}} \bibnamefont{and}
  \bibinfo{author}{\bibfnamefont{H.~A.} \bibnamefont{Bethe}},
  \bibinfo{journal}{Phys. Rev.} \textbf{\bibinfo{volume}{84}},
  \bibinfo{pages}{1232} (\bibinfo{year}{1951}).

\bibitem[{\citenamefont{Hellstern et~al.}(1997)\citenamefont{Hellstern,
  Alkofer, Oettel, and Reinhardt}}]{HAOR97}
\bibinfo{author}{\bibfnamefont{G.}~\bibnamefont{Hellstern}},
  \bibinfo{author}{\bibfnamefont{R.}~\bibnamefont{Alkofer}},
  \bibinfo{author}{\bibfnamefont{M.}~\bibnamefont{Oettel}}, \bibnamefont{and}
  \bibinfo{author}{\bibfnamefont{H.}~\bibnamefont{Reinhardt}},
  \bibinfo{journal}{Nucl. Phys.} \textbf{\bibinfo{volume}{A627}},
  \bibinfo{pages}{679} (\bibinfo{year}{1997}),
  \eprint[http://arXiv.org/abs]{hep-ph/9705267}.

\bibitem[{\citenamefont{Oettel et~al.}(1998)\citenamefont{Oettel, Hellstern,
  Alkofer, and Reinhardt}}]{OHAR98}
\bibinfo{author}{\bibfnamefont{M.}~\bibnamefont{Oettel}},
  \bibinfo{author}{\bibfnamefont{G.}~\bibnamefont{Hellstern}},
  \bibinfo{author}{\bibfnamefont{R.}~\bibnamefont{Alkofer}}, \bibnamefont{and}
  \bibinfo{author}{\bibfnamefont{H.}~\bibnamefont{Reinhardt}},
  \bibinfo{journal}{Phys. Rev.} \textbf{\bibinfo{volume}{C58}},
  \bibinfo{pages}{2459} (\bibinfo{year}{1998}),
  \eprint[http://arXiv.org/abs]{nucl-th/9805054}.

\bibitem[{\citenamefont{Walecka}(1974)}]{Walecka74}
\bibinfo{author}{\bibfnamefont{J.~D.} \bibnamefont{Walecka}},
  \bibinfo{journal}{Annals Phys.} \textbf{\bibinfo{volume}{83}},
  \bibinfo{pages}{491} (\bibinfo{year}{1974}).

\bibitem[{\citenamefont{Serot and Walecka}(1986)}]{SW86}
\bibinfo{author}{\bibfnamefont{B.~D.} \bibnamefont{Serot}} \bibnamefont{and}
  \bibinfo{author}{\bibfnamefont{J.~D.} \bibnamefont{Walecka}},
  \bibinfo{journal}{Adv. Nucl. Phys.} \textbf{\bibinfo{volume}{16}},
  \bibinfo{pages}{1} (\bibinfo{year}{1986}).

\bibitem[{\citenamefont{Ichinose et~al.}(2001)\citenamefont{Ichinose, Matsui,
  and Onoda}}]{IMO01}
\bibinfo{author}{\bibfnamefont{I.}~\bibnamefont{Ichinose}},
  \bibinfo{author}{\bibfnamefont{T.}~\bibnamefont{Matsui}}, \bibnamefont{and}
  \bibinfo{author}{\bibfnamefont{M.}~\bibnamefont{Onoda}},
  \bibinfo{journal}{Phys. Rev.} \textbf{\bibinfo{volume}{B64}},
  \bibinfo{pages}{104516} (\bibinfo{year}{2001}).

\bibitem[{\citenamefont{Cahill}(1989)}]{Cahill89}
\bibinfo{author}{\bibfnamefont{R.}~\bibnamefont{Cahill}},
  \bibinfo{journal}{Aust. J. Phys.} \textbf{\bibinfo{volume}{42}},
  \bibinfo{pages}{171} (\bibinfo{year}{1989}).

\bibitem[{\citenamefont{Reinhardt}(1990)}]{Rein90}
\bibinfo{author}{\bibfnamefont{H.}~\bibnamefont{Reinhardt}},
  \bibinfo{journal}{Phys. Lett.} \textbf{\bibinfo{volume}{B244}},
  \bibinfo{pages}{316} (\bibinfo{year}{1990}).

\bibitem[{\citenamefont{Ebert and Kaschluhn}(1992)}]{Ebka92}
\bibinfo{author}{\bibfnamefont{D.}~\bibnamefont{Ebert}} \bibnamefont{and}
  \bibinfo{author}{\bibfnamefont{L.}~\bibnamefont{Kaschluhn}},
  \bibinfo{journal}{Phys. Lett.} \textbf{\bibinfo{volume}{B297}},
  \bibinfo{pages}{367} (\bibinfo{year}{1992}).

\bibitem[{\citenamefont{Ebert et~al.}(1996)\citenamefont{Ebert, Feldmann,
  Kettner, and Reinhardt}}]{EFKR96}
\bibinfo{author}{\bibfnamefont{D.}~\bibnamefont{Ebert}},
  \bibinfo{author}{\bibfnamefont{T.}~\bibnamefont{Feldmann}},
  \bibinfo{author}{\bibfnamefont{C.}~\bibnamefont{Kettner}}, \bibnamefont{and}
  \bibinfo{author}{\bibfnamefont{H.}~\bibnamefont{Reinhardt}},
  \bibinfo{journal}{Z. Phys.} \textbf{\bibinfo{volume}{C71}},
  \bibinfo{pages}{329} (\bibinfo{year}{1996}),
  \eprint[http://arXiv.org/abs]{hep-ph/9506298}.

\bibitem[{\citenamefont{Ebert et~al.}(1998)\citenamefont{Ebert, Feldmann,
  Kettner, and Reinhardt}}]{EFKR98}
\bibinfo{author}{\bibfnamefont{D.}~\bibnamefont{Ebert}},
  \bibinfo{author}{\bibfnamefont{T.}~\bibnamefont{Feldmann}},
  \bibinfo{author}{\bibfnamefont{C.}~\bibnamefont{Kettner}}, \bibnamefont{and}
  \bibinfo{author}{\bibfnamefont{H.}~\bibnamefont{Reinhardt}},
  \bibinfo{journal}{Int. J. Mod. Phys.} \textbf{\bibinfo{volume}{A13}},
  \bibinfo{pages}{1091} (\bibinfo{year}{1998}),
  \eprint[http://arXiv.org/abs]{hep-ph/9601257}.

\bibitem[{\citenamefont{Ebert and Jurke}(1998)}]{EJ98}
\bibinfo{author}{\bibfnamefont{D.}~\bibnamefont{Ebert}} \bibnamefont{and}
  \bibinfo{author}{\bibfnamefont{T.}~\bibnamefont{Jurke}},
  \bibinfo{journal}{Phys. Rev.} \textbf{\bibinfo{volume}{D58}},
  \bibinfo{pages}{034001} (\bibinfo{year}{1998}),
  \eprint[http://arXiv.org/abs]{hep-ph/9710390}.

\bibitem[{\citenamefont{Cahill et~al.}(1987)\citenamefont{Cahill, Roberts, and
  Praschifka}}]{CRCP87}
\bibinfo{author}{\bibfnamefont{R.~T.} \bibnamefont{Cahill}},
  \bibinfo{author}{\bibfnamefont{C.~D.} \bibnamefont{Roberts}},
  \bibnamefont{and}
  \bibinfo{author}{\bibfnamefont{J.}~\bibnamefont{Praschifka}},
  \bibinfo{journal}{Phys. Rev.} \textbf{\bibinfo{volume}{D36}},
  \bibinfo{pages}{2804} (\bibinfo{year}{1987}).

\bibitem[{\citenamefont{Vogl}(1990)}]{Vogl90}
\bibinfo{author}{\bibfnamefont{U.}~\bibnamefont{Vogl}}, \bibinfo{journal}{Z.
  Phys.} \textbf{\bibinfo{volume}{A337}}, \bibinfo{pages}{191}
  (\bibinfo{year}{1990}).

\bibitem[{\citenamefont{Mineo et~al.}(2002)\citenamefont{Mineo, Bentz, Ishii,
  and Yazaki}}]{MBIY02}
\bibinfo{author}{\bibfnamefont{H.}~\bibnamefont{Mineo}},
  \bibinfo{author}{\bibfnamefont{W.}~\bibnamefont{Bentz}},
  \bibinfo{author}{\bibfnamefont{N.}~\bibnamefont{Ishii}}, \bibnamefont{and}
  \bibinfo{author}{\bibfnamefont{K.}~\bibnamefont{Yazaki}}
  (\bibinfo{year}{2002}), \eprint[http://arXiv.org/abs]{nucl-th/0201082}.

\bibitem[{\citenamefont{Hatsuda and Kunihiro}(1994)}]{HK94}
\bibinfo{author}{\bibfnamefont{T.}~\bibnamefont{Hatsuda}} \bibnamefont{and}
  \bibinfo{author}{\bibfnamefont{T.}~\bibnamefont{Kunihiro}},
  \bibinfo{journal}{Phys. Rept.} \textbf{\bibinfo{volume}{247}},
  \bibinfo{pages}{221} (\bibinfo{year}{1994}), \bibinfo{note}{and references
  therein}, \eprint[http://arXiv.org/abs]{hep-ph/9401310}.

\bibitem[{\citenamefont{Ebert et~al.}(1994)\citenamefont{Ebert, Reinhardt, and
  Volkov}}]{ERV94}
\bibinfo{author}{\bibfnamefont{D.}~\bibnamefont{Ebert}},
  \bibinfo{author}{\bibfnamefont{H.}~\bibnamefont{Reinhardt}},
  \bibnamefont{and} \bibinfo{author}{\bibfnamefont{M.~K.}
  \bibnamefont{Volkov}}, \bibinfo{journal}{Prog. Part. Nucl. Phys.}
  \textbf{\bibinfo{volume}{33}}, \bibinfo{pages}{1} (\bibinfo{year}{1994}).

\bibitem[{\citenamefont{Kawamoto and Smit}(1981)}]{KS81}
\bibinfo{author}{\bibfnamefont{N.}~\bibnamefont{Kawamoto}} \bibnamefont{and}
  \bibinfo{author}{\bibfnamefont{J.}~\bibnamefont{Smit}},
  \bibinfo{journal}{Nucl. Phys.} \textbf{\bibinfo{volume}{B192}},
  \bibinfo{pages}{100} (\bibinfo{year}{1981}).

\bibitem[{\citenamefont{Andrianov et~al.}(1987)\citenamefont{Andrianov,
  Andrianov, Novozhilov, and Novozhilov}}]{And87}
\bibinfo{author}{\bibfnamefont{A.~A.} \bibnamefont{Andrianov}},
  \bibinfo{author}{\bibfnamefont{V.~A.} \bibnamefont{Andrianov}},
  \bibinfo{author}{\bibfnamefont{V.~Y.} \bibnamefont{Novozhilov}},
  \bibnamefont{and} \bibinfo{author}{\bibfnamefont{Y.~V.}
  \bibnamefont{Novozhilov}}, \bibinfo{journal}{Phys. Lett.}
  \textbf{\bibinfo{volume}{B186}}, \bibinfo{pages}{401} (\bibinfo{year}{1987}).

\bibitem[{\citenamefont{Ball}(1990)}]{Ball90}
\bibinfo{author}{\bibfnamefont{R.~D.} \bibnamefont{Ball}},
  \bibinfo{journal}{Int. J. Mod. Phys.} \textbf{\bibinfo{volume}{A5}},
  \bibinfo{pages}{4391} (\bibinfo{year}{1990}).

\bibitem[{\citenamefont{Reinhardt}(1991)}]{Rein91}
\bibinfo{author}{\bibfnamefont{H.}~\bibnamefont{Reinhardt}},
  \bibinfo{journal}{Phys. Lett.} \textbf{\bibinfo{volume}{B257}},
  \bibinfo{pages}{375} (\bibinfo{year}{1991}).

\bibitem[{\citenamefont{Bijnens et~al.}(1993)\citenamefont{Bijnens, Bruno, and
  de~Rafael}}]{BBR93}
\bibinfo{author}{\bibfnamefont{J.}~\bibnamefont{Bijnens}},
  \bibinfo{author}{\bibfnamefont{C.}~\bibnamefont{Bruno}}, \bibnamefont{and}
  \bibinfo{author}{\bibfnamefont{E.}~\bibnamefont{de~Rafael}},
  \bibinfo{journal}{Nucl. Phys.} \textbf{\bibinfo{volume}{B390}},
  \bibinfo{pages}{501} (\bibinfo{year}{1993}),
  \eprint[http://arXiv.org/abs]{hep-ph/9206236}.

\bibitem[{\citenamefont{Schaden et~al.}(1990)\citenamefont{Schaden, Reinhardt,
  Amundsen, and Lavelle}}]{SRAL}
\bibinfo{author}{\bibfnamefont{M.}~\bibnamefont{Schaden}},
  \bibinfo{author}{\bibfnamefont{H.}~\bibnamefont{Reinhardt}},
  \bibinfo{author}{\bibfnamefont{P.~A.} \bibnamefont{Amundsen}},
  \bibnamefont{and} \bibinfo{author}{\bibfnamefont{M.~J.}
  \bibnamefont{Lavelle}}, \bibinfo{journal}{Nucl. Phys.}
  \textbf{\bibinfo{volume}{B339}}, \bibinfo{pages}{595} (\bibinfo{year}{1990}).

\bibitem[{\citenamefont{Gockeler et~al.}(1990)}]{Gock90}
\bibinfo{author}{\bibfnamefont{M.}~\bibnamefont{Gockeler}}
  \bibnamefont{et~al.}, \bibinfo{journal}{Nucl. Phys.}
  \textbf{\bibinfo{volume}{B334}}, \bibinfo{pages}{527} (\bibinfo{year}{1990}).

\bibitem[{\citenamefont{Belkov et~al.}(1993)\citenamefont{Belkov, Ebert, and
  Emelyanenko}}]{BEE93}
\bibinfo{author}{\bibfnamefont{A.~A.} \bibnamefont{Belkov}},
  \bibinfo{author}{\bibfnamefont{D.}~\bibnamefont{Ebert}}, \bibnamefont{and}
  \bibinfo{author}{\bibfnamefont{A.~V.} \bibnamefont{Emelyanenko}},
  \bibinfo{journal}{Nucl. Phys.} \textbf{\bibinfo{volume}{A552}},
  \bibinfo{pages}{523} (\bibinfo{year}{1993}).

\bibitem[{\citenamefont{Stratonovich}(1958)}]{Strat58}
\bibinfo{author}{\bibfnamefont{R.~L.} \bibnamefont{Stratonovich}},
  \bibinfo{journal}{Sov. Phys. Dokl.} \textbf{\bibinfo{volume}{2}},
  \bibinfo{pages}{416} (\bibinfo{year}{1958}).

\bibitem[{\citenamefont{Hubbard}(1959)}]{Hubbard59}
\bibinfo{author}{\bibfnamefont{J.}~\bibnamefont{Hubbard}},
  \bibinfo{journal}{Phys. Rev. Lett.} \textbf{\bibinfo{volume}{3}},
  \bibinfo{pages}{77} (\bibinfo{year}{1959}).

\bibitem[{\citenamefont{Espriu et~al.}(1983)\citenamefont{Espriu, Pascual, and
  Tarrach}}]{EPT83}
\bibinfo{author}{\bibfnamefont{D.}~\bibnamefont{Espriu}},
  \bibinfo{author}{\bibfnamefont{P.}~\bibnamefont{Pascual}}, \bibnamefont{and}
  \bibinfo{author}{\bibfnamefont{R.}~\bibnamefont{Tarrach}},
  \bibinfo{journal}{Nucl. Phys.} \textbf{\bibinfo{volume}{B214}},
  \bibinfo{pages}{285} (\bibinfo{year}{1983}).

\bibitem[{\citenamefont{Greiner and Schaefer}(1994)}]{Greiner94}
\bibinfo{author}{\bibfnamefont{W.}~\bibnamefont{Greiner}} \bibnamefont{and}
  \bibinfo{author}{\bibfnamefont{A.}~\bibnamefont{Schaefer}},
  \emph{\bibinfo{title}{Quantum chromodynamics}}
  (\bibinfo{publisher}{Springer}, \bibinfo{address}{Berlin, Germany},
  \bibinfo{year}{1994}), \bibinfo{note}{p 414}.

\bibitem[{\citenamefont{Hosaka and Toki}(2001)}]{Hosaka01}
\bibinfo{author}{\bibfnamefont{A.}~\bibnamefont{Hosaka}} \bibnamefont{and}
  \bibinfo{author}{\bibfnamefont{H.}~\bibnamefont{Toki}},
  \emph{\bibinfo{title}{Quarks, baryons and chiral symmetry}}
  (\bibinfo{publisher}{World Scientific}, \bibinfo{address}{Singapore},
  \bibinfo{year}{2001}), \bibinfo{note}{p 88}.

\bibitem[{\citenamefont{Dumbrajs et~al.}(1983)}]{Dumbrajs83}
\bibinfo{author}{\bibfnamefont{O.}~\bibnamefont{Dumbrajs}}
  \bibnamefont{et~al.}, \bibinfo{journal}{Nucl. Phys.}
  \textbf{\bibinfo{volume}{B216}}, \bibinfo{pages}{277} (\bibinfo{year}{1983}).

\bibitem[{\citenamefont{Groom et~al.}(2000)}]{particle00}
\bibinfo{author}{\bibfnamefont{D.~E.} \bibnamefont{Groom}} \bibnamefont{et~al.}
  (\bibinfo{collaboration}{Particle Data Group}), \bibinfo{journal}{Eur. Phys.
  J.} \textbf{\bibinfo{volume}{C15}}, \bibinfo{pages}{1}
  (\bibinfo{year}{2000}).

\bibitem[{\citenamefont{Dziembowski et~al.}(1981)\citenamefont{Dziembowski,
  Metzger, and Van~de Walle}}]{DMW81}
\bibinfo{author}{\bibfnamefont{Z.}~\bibnamefont{Dziembowski}},
  \bibinfo{author}{\bibfnamefont{W.~J.} \bibnamefont{Metzger}},
  \bibnamefont{and} \bibinfo{author}{\bibfnamefont{R.~T.} \bibnamefont{Van~de
  Walle}}, \bibinfo{journal}{Z. Phys.} \textbf{\bibinfo{volume}{C10}},
  \bibinfo{pages}{231} (\bibinfo{year}{1981}).

\bibitem[{\citenamefont{H\"ohler et~al.}(1976)}]{Hohler76}
\bibinfo{author}{\bibfnamefont{G.}~\bibnamefont{H\"ohler}}
  \bibnamefont{et~al.}, \bibinfo{journal}{Nucl. Phys.}
  \textbf{\bibinfo{volume}{B114}}, \bibinfo{pages}{505} (\bibinfo{year}{1976}).

\bibitem[{\citenamefont{Weiss et~al.}(1993)\citenamefont{Weiss, Buck, Alkofer,
  and Reinhardt}}]{WBAR93}
\bibinfo{author}{\bibfnamefont{C.}~\bibnamefont{Weiss}},
  \bibinfo{author}{\bibfnamefont{A.}~\bibnamefont{Buck}},
  \bibinfo{author}{\bibfnamefont{R.}~\bibnamefont{Alkofer}}, \bibnamefont{and}
  \bibinfo{author}{\bibfnamefont{H.}~\bibnamefont{Reinhardt}},
  \bibinfo{journal}{Phys. Lett.} \textbf{\bibinfo{volume}{B312}},
  \bibinfo{pages}{6} (\bibinfo{year}{1993}),
  \eprint[http://arXiv.org/abs]{hep-ph/9305215}.

\bibitem[{\citenamefont{Eden et~al.}(1994)}]{Eden94}
\bibinfo{author}{\bibfnamefont{T.}~\bibnamefont{Eden}} \bibnamefont{et~al.},
  \bibinfo{journal}{Phys. Rev.} \textbf{\bibinfo{volume}{C50}},
  \bibinfo{pages}{1749} (\bibinfo{year}{1994}).

\bibitem[{\citenamefont{Bruins et~al.}(1995)}]{Bruins95}
\bibinfo{author}{\bibfnamefont{E.~E.~W.} \bibnamefont{Bruins}}
  \bibnamefont{et~al.}, \bibinfo{journal}{Phys. Rev. Lett.}
  \textbf{\bibinfo{volume}{75}}, \bibinfo{pages}{21} (\bibinfo{year}{1995}).

\bibitem[{\citenamefont{Ostrick et~al.}(1999)}]{Ostrick99}
\bibinfo{author}{\bibfnamefont{M.}~\bibnamefont{Ostrick}} \bibnamefont{et~al.},
  \bibinfo{journal}{Phys. Rev. Lett.} \textbf{\bibinfo{volume}{83}},
  \bibinfo{pages}{276} (\bibinfo{year}{1999}).

\bibitem[{\citenamefont{Rohe et~al.}(1999)}]{Rohe99}
\bibinfo{author}{\bibfnamefont{D.}~\bibnamefont{Rohe}} \bibnamefont{et~al.},
  \bibinfo{journal}{Phys. Rev. Lett.} \textbf{\bibinfo{volume}{83}},
  \bibinfo{pages}{4257} (\bibinfo{year}{1999}).

\bibitem[{\citenamefont{Zhu et~al.}(2001)}]{Zhu01}
\bibinfo{author}{\bibfnamefont{H.}~\bibnamefont{Zhu}} \bibnamefont{et~al.}
  (\bibinfo{collaboration}{E93026}), \bibinfo{journal}{Phys. Rev. Lett.}
  \textbf{\bibinfo{volume}{87}}, \bibinfo{pages}{081801}
  (\bibinfo{year}{2001}), \eprint[http://arXiv.org/abs]{nucl-ex/0105001}.

\bibitem[{\citenamefont{Anklin et~al.}(1994)}]{Anklin94}
\bibinfo{author}{\bibfnamefont{H.}~\bibnamefont{Anklin}} \bibnamefont{et~al.},
  \bibinfo{journal}{Phys. Lett.} \textbf{\bibinfo{volume}{B336}},
  \bibinfo{pages}{313} (\bibinfo{year}{1994}).

\bibitem[{\citenamefont{Anklin et~al.}(1998)}]{Anklin98}
\bibinfo{author}{\bibfnamefont{H.}~\bibnamefont{Anklin}} \bibnamefont{et~al.},
  \bibinfo{journal}{Phys. Lett.} \textbf{\bibinfo{volume}{B428}},
  \bibinfo{pages}{248} (\bibinfo{year}{1998}).

\bibitem[{\citenamefont{Golak et~al.}(2001)\citenamefont{Golak, Ziemer, Kamada,
  Witala, and Glockle}}]{Golak00}
\bibinfo{author}{\bibfnamefont{J.}~\bibnamefont{Golak}},
  \bibinfo{author}{\bibfnamefont{G.}~\bibnamefont{Ziemer}},
  \bibinfo{author}{\bibfnamefont{H.}~\bibnamefont{Kamada}},
  \bibinfo{author}{\bibfnamefont{H.}~\bibnamefont{Witala}}, \bibnamefont{and}
  \bibinfo{author}{\bibfnamefont{W.}~\bibnamefont{Glockle}},
  \bibinfo{journal}{Phys. Rev.} \textbf{\bibinfo{volume}{C63}},
  \bibinfo{pages}{034006} (\bibinfo{year}{2001}),
  \eprint[http://arXiv.org/abs]{nucl-th/0008008}.

\bibitem[{\citenamefont{Xu et~al.}(2000)}]{Xu00}
\bibinfo{author}{\bibfnamefont{W.}~\bibnamefont{Xu}} \bibnamefont{et~al.},
  \bibinfo{journal}{Phys. Rev. Lett.} \textbf{\bibinfo{volume}{85}},
  \bibinfo{pages}{2900} (\bibinfo{year}{2000}),
  \eprint[http://arXiv.org/abs]{nucl-ex/0008003}.

\bibitem[{\citenamefont{Kubon et~al.}(2002)}]{Kubon01}
\bibinfo{author}{\bibfnamefont{G.}~\bibnamefont{Kubon}} \bibnamefont{et~al.},
  \bibinfo{journal}{Phys. Lett.} \textbf{\bibinfo{volume}{B524}},
  \bibinfo{pages}{26} (\bibinfo{year}{2002}),
  \eprint[http://arXiv.org/abs]{nucl-ex/0107016}.

\end{thebibliography}
\end{document}